\pdfoutput=1 
\documentclass[dvipsnames]{elsarticle}
\usepackage{hyperref}
\usepackage{graphicx}  
\usepackage{dcolumn}   
\usepackage{bm}        
\usepackage{amssymb}   
\usepackage{amsfonts}
\usepackage{graphics}
\usepackage{grffile}   
\usepackage{epsfig}
\usepackage[usenames]{color}
\usepackage[normalem]{ulem} 
\usepackage{color}
\usepackage[utf8]{inputenc}
\usepackage[T1]{fontenc}
\usepackage{amsmath}
\usepackage{booktabs}
\usepackage{svg}
\usepackage{dsfont}

\usepackage{xcolor}
\usepackage{tikz}
\usepackage{environ}
\makeatletter
    \newsavebox{\measure@tikzpicture}
        \NewEnviron{scaletikzpicturetowidth}[1]{%
            \def\tikz@width{#1}%
            \def\tikzscale{1}\begin{lrbox}{\measure@tikzpicture}%
            \BODY
            \end{lrbox}%
            \pgfmathparse{#1/\wd\measure@tikzpicture}%
            \edef\tikzscale{\pgfmathresult}%
            \BODY
        }
\makeatother
\usetikzlibrary{math}
\usetikzlibrary{calc}
\usetikzlibrary{decorations.pathreplacing, decorations.markings}
\usetikzlibrary{patterns, arrows, arrows.meta}

\usepackage{subcaption}
\usepackage{siunitx}

\DeclareSIUnit{\ele}{\mbox{$e^{\text{-}}$}}
\DeclareSIUnit[number-unit-product = ]{\percent}{\%}
\DeclareSIUnit{\raddose}{rad}
\DeclareSIUnit{\tid}{\mbox{\si{\kilo\raddose}}}
\DeclareSIUnit{\niel}{\mbox{\SI{1}{\mega\electronvolt}\ n$_{\mathrm{eq}}$\ \si{\per\cm\squared}}}
\DeclareSIUnit{\pixel}{pixel}

\newcommand{\Vsub}{V_{sub}}
\newcommand{\Vpwell}{V_{pwell}}
\newcommand{\Vh}{V_{h}}
\newcommand{\Vb}{V_{casb}}
\newcommand{\Vn}{V_{casn}}
\newcommand{\Ir}{I_{reset}}
\newcommand{\Id}{I_{db}}
\newcommand{\Ib}{I_{bias}}
\newcommand{\Ibn}{I_{biasn}}

\journal{Nucl. Instrum. Methods Phys. Res. A}
\begin{document}%
\begin{frontmatter}
\title{Digital Pixel Test Structures implemented in a 65~nm CMOS process}

\author[1]{Gianluca Aglieri Rinella}
\author[13]{Anton Andronic}
\author[6]{Matias Antonelli}
\author[9,10]{Mauro Aresti}
\author[6]{Roberto Baccomi}
\author[2]{Pascal Becht}
\author[3,4]{Stefania Beole}
\author[1,18]{Justus Braach}
\author[5,6]{Matthew Daniel Buckland}
\author[1]{Eric Buschmann}
\author[5,6]{Paolo Camerini}
\author[1]{Francesca Carnesecchi}
\author[1]{Leonardo Cecconi}
\author[17]{Edoardo Charbon}
\author[5,6]{Giacomo Contin}
\author[1]{Dominik Dannheim}
\author[1]{Joao de Melo}
\author[1,15]{Wenjing Deng}
\author[1]{Antonello di Mauro}
\author[1]{Jan Hasenbichler}
\author[1]{Hartmut Hillemanns}
\author[1,16]{Geun Hee Hong}
\author[7]{Artem Isakov}
\author[1]{Antoine Junique}
\author[1]{Alex Kluge}
\author[7]{Artem Kotliarov}
\author[7]{Filip K\v{r}\'i\v{z}ek}
\author[1,8]{Lukas Lautner}
\author[1]{Magnus Mager}
\author[9,10]{Davide Marras}
\author[1]{Paolo Martinengo}
\author[2]{Silvia Masciocchi}
\author[2]{Marius Wilm Menzel}
\author[1]{Magdalena Munker}
\author[1,17]{Francesco Piro}
\author[6]{Alexandre Rachevski}
\author[1]{Karoliina Rebane}
\author[1]{Felix Reidt}
\author[11]{Roberto Russo}
\author[1]{Isabella Sanna}
\author[9,10]{Valerio Sarritzu}
\author[12]{Serhiy Senyukov}
\author[1]{Walter Snoeys}
\author[11]{Jory Sonneveld}
\author[1]{Miljenko \v{S}ulji\'c\corref{cor1}}
\author[1]{Peter Svihra}
\author[13]{Nicolas Tiltmann}
\author[9,10]{Gianluca Usai}
\author[1]{Jacob Bastiaan Van Beelen}
\author[14,19]{Mirella Dimitrova Vassilev}
\author[14,19]{Caterina Vernieri}
\author[5,6]{Anna Villani}

\affiliation[1]{organization={European Organisation for Nuclear Research (CERN)}, city={Geneva}, country={Switzerland}}
\affiliation[13]{organization={Westf\"alische Wilhelms-Universit\"at M\"unster}, city={M\"unster}, country={Germany}}
\affiliation[2]{organization={Ruprecht Karls Universit\"at Heidelberg}, city={Heidelberg}, country={Germany}}
\affiliation[3]{organization={Universit\`a degli studi di Torino}, city={Torino}, country={Italy}}
\affiliation[4]{organization={Istituto Nazionale di Fisica Nucleare (INFN), Sezione di Torino}, city={Torino}, country={Italy}}
\affiliation[18]{organization={Universit\"at Hamburg}, city={Hamburg}, country={Germany}}
\affiliation[5]{organization={Universit\`a degli studi di Trieste}, city={Trieste}, country={Italy}}
\affiliation[6]{organization={Istituto Nazionale di Fisica Nucleare (INFN), Sezione di Trieste}, city={Trieste}, country={Italy}}
\affiliation[17]{organization={EPFL}, city={Lausanne}, country={Switzerland}}
\affiliation[15]{organization={Central China Normal University (CCNU)}, city={Wuhan}, country={China}}
\affiliation[16]{organization={Yonsei University}, city={Seoul}, country={Republic of Korea}}
\affiliation[7]{organization={Czech Academy of Sciences}, city={Prague}, country={Czechia}}
\affiliation[8]{organization={Technische Universit\"at M\"unchen}, city={M\"unchen}, country={Germany}}
\affiliation[9]{organization={Universit\`a degli studi di Cagliari}, city={Cagliari}, country={Italy}}
\affiliation[10]{organization={Istituto Nazionale di Fisica Nucleare (INFN), Sezione di Cagliari}, city={Cagliari}, country={Italy}}
\affiliation[11]{organization={Nikhef National institute for subatomic physics}, city={Amsterdam}, country={Netherlands}}
\affiliation[12]{organization={Centre National de la Recherche Scientifique}, city={Strasbourg}, country={France}}
\affiliation[14]{organization={SLAC}, city={Menlo Park}, country={California, USA}}
\affiliation[19]{organization={Stanford University}, city={Stanford}, country={California, USA}}

\cortext[cor1]{Corresponding author}

\begin{abstract}

The ALICE ITS3 (Inner Tracking System 3) upgrade project and the CERN EP R\&D on monolithic pixel sensors are investigating the feasibility of the Tower Partners Semiconductor Co.\ \SI{65}{\nm} process for use in the next generation of vertex detectors. The ITS3 aims to employ wafer-scale Monolithic Active Pixel Sensors thinned down to \SIrange{20}{40}{\um} and bent to form truly cylindrical half barrels. Among the first critical steps towards the realisation of this detector is to validate the sensor technology through extensive characterisation both in the laboratory and with in-beam measurements. The Digital Pixel Test Structure (DPTS) is one of the prototypes produced in the first sensor submission in this technology and has undergone a systematic measurement campaign whose details are presented in this article. 

The results confirm the goals of detection efficiency and non-ionising and ionising radiation hardness up to the expected levels for ALICE ITS3 and also demonstrate operation at +\SI{20}{\celsius} and a detection efficiency of \SI{99}{\percent} for a DPTS irradiated with a dose of \SI{e15}{\niel}.
Furthermore, spatial, timing and energy resolutions were measured at various settings and irradiation levels.

\end{abstract}

\begin{keyword}
Monolithic Active Pixel Sensors \sep Solid state detectors
\end{keyword}

\end{frontmatter}

%
%





\tikzset{
	beamarrow/.style={
		decoration={
			markings,mark=at position 1 with 
			{\arrow[scale=2,>=stealth]{>}}
		},postaction={decorate}
	}
}
\tikzset{
	pics/.cd,
	vector out/.style={
		code={
		\draw[#1, thick] (0,0)  circle (0.15) (45:0.15) -- (225:0.15) (135:0.15) -- (315:0.15);
		}
	}
}
\tikzset{
	pics/.cd,
	vector in/.style={
		code={
		\draw[#1, thick] (0,0)  circle (0.15);
		 \fill[#1] (0,0)  circle (.05);
		 }
	}
}
\tikzset{
	global scale/.style={
		scale=#1,
		every node/.style={scale=#1}
	}
}
\def\centerarc[#1] (#2)(#3:#4:#5) 
	 { \draw[#1] ($(#2)+({#5*cos(#3)},{#5*sin(#3)})$) arc (#3:#4:#5); }
\section{Introduction}
\label{sec:intro}

Over the past decade, Monolithic Active Pixel Sensors (MAPS) have well established their position in high energy physics experiment vertex detectors~\cite{ls2paper,its2conf,star}. By combining the readout circuitry and the sensitive volume in the same sensor produced in commercial processes, they pave the way for ultra-thin and large-scale tracking detectors.
The state-of-art ALPIDE chip, in use in the \mbox{ALICE} Inner Tracking System 2 (ITS2), was produced in the TowerJazz \qty{180}{\nm}~technology and demonstrated an excellent detection efficiency (\qty{\gg 99}{\percent}) and spatial resolution (\qty{5}{\um}) performance at very low power dissipation (\qty{<40}{\milli\watt\per\cm^2}) and material budget of \qty{0.05}{\percent X/X_0} up to irradiation levels of about \qty{3}{\kilo\gray} and \qty{e13}{\niel} \cite{ALPIDE-proceedings-1, ALPIDE-proceedings-2, ALPIDE-proceedings-3}.
Other developments produced in the same technology demonstrated comparable detection efficiency at \qty{-20}{\degreeCelsius} even after a non-ionising radiation dose of \qty{e15}{\niel} \cite{malta2}.
The next upgrade of ALICE tracker, the ALICE ITS3, planned for the Long Shutdowns 3 (2026-2028), is extending the requirements on the sensor characteristics.
The ITS3 detector integration imposes a wafer-scale sensor size $O(10\times\qty{27}{\cm^2})$ and operation at temperatures achievable by air cooling $O(\qty{+20}{\degreeCelsius})$ \cite{loi}.

In order to address the arising challenges in the design of the next generation MAPS detector, the Tower Partners Semiconductor Co.~(TPSCo) \SI{65}{\nm} CMOS imaging process~\cite{tower} was chosen as the starting point in the framework of the ALICE~ITS3 upgrade~\cite{loi,its3conf} and the CERN~EP~R\&D on monolithic pixel sensors~\cite{eprnd}. The critical aspect of the new technology node for the ALICE~ITS3 is the larger wafer size (\qty{30}{\cm} as opposed to \qty{20}{\cm} in TowerJazz \qty{180}{\nm}~process). The goal of the first submission designated MLR1 and produced in summer~2021, was to verify the radiation hardness (the expected levels for the ITS3 are below \qty{e13}{\niel} and \qty{10}{\kilo\gray} \cite{loi}) and the detection efficiency ($>$\qty{99}{\percent}) of the MAPS produced in this technology.
Therefore, a sensor prototype featuring in-pixel amplification and discrimination was characterised using soft X-rays and ionising particle beams and the results are described in this paper. In particular, the sensors were irradiated\footnote{Sensors irradiated with non-ionising, ionising, and combined doses were exposed to neutrons at JSI Ljubljana, \qty{10}{\keV} X-rays from a tungsten target at CERN, and \qty{30}{\MeV} protons at NPI Prague, respectively.} up to \qty{100}{\kilo\gray} and \qty{e15}{\niel} and their detection efficiency and spatial, timing and energy resolutions were measured. In total, \num{48} sensors have been tested, with at least two for each radiation level (three in case of \qty{e15}{\niel}). For visualisation purposes and to eliminate any selection bias, each measurement point is represented by a single device, chosen as the first one that was tested.

\section{The DPTS chip}
\label{sec:dpts_chip}

\begin{figure}[!htb]
    \begin{subfigure}[t]{0.6\textwidth}
	    \includegraphics[width=\textwidth]{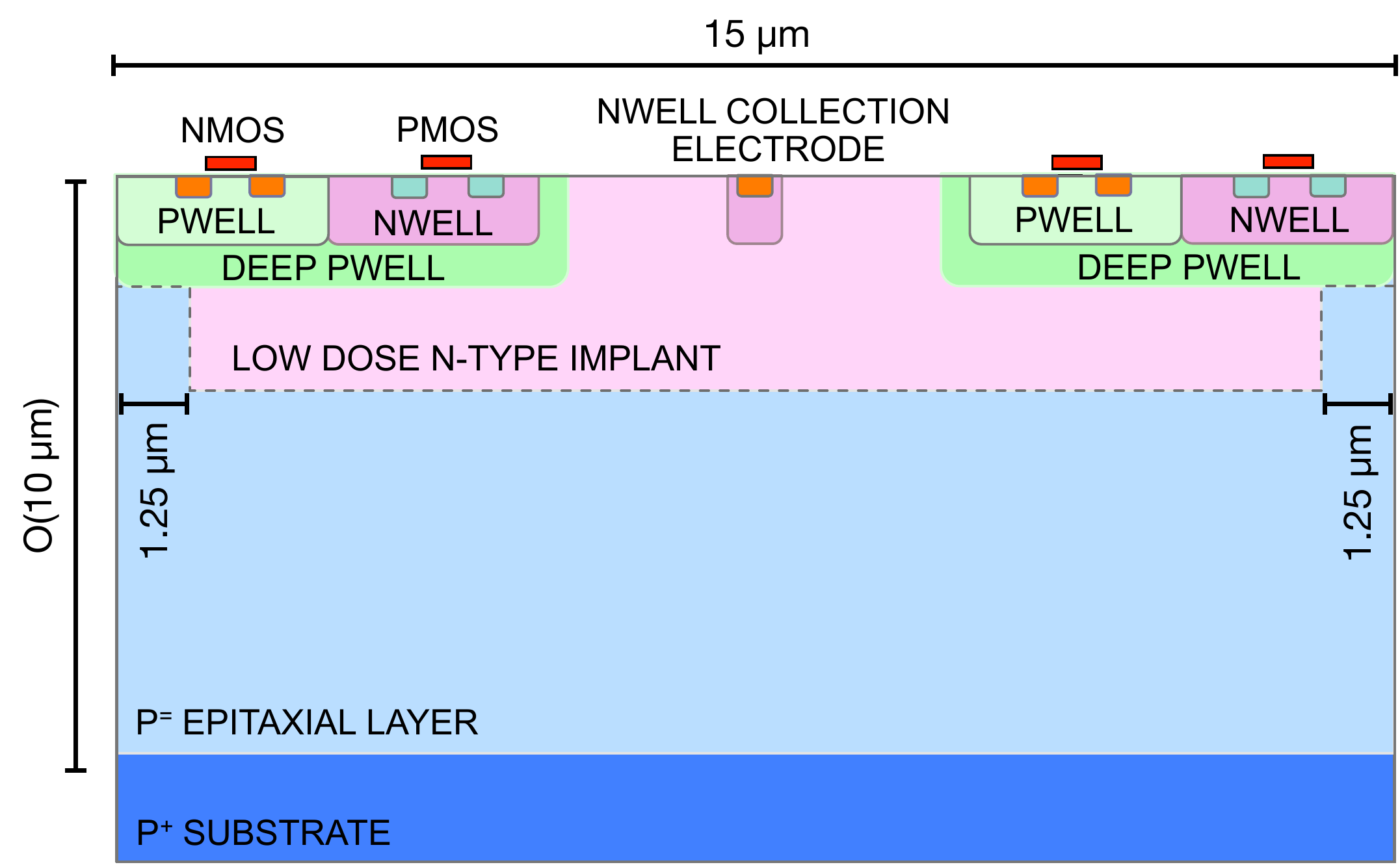}
	    \caption{Pixel cross section. Not to scale.}
        \label{fig:dpts:cross_section}
	\end{subfigure}
    \hspace{0.5mm}
    \begin{subfigure}[t]{0.37\textwidth}
        \includegraphics[width=\textwidth]{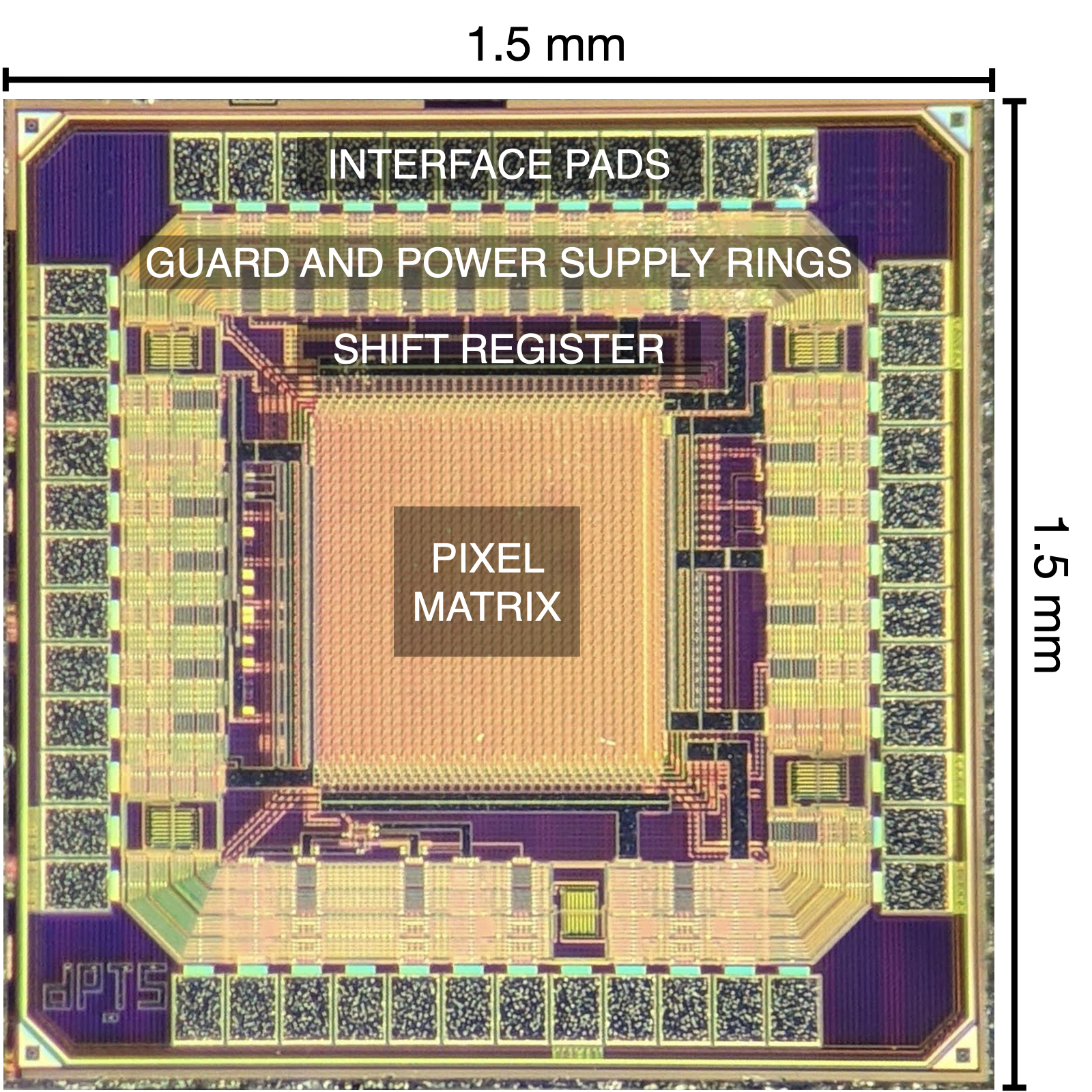}
        \caption{Chip under microscope.}
        \label{fig:dpts:photo}
    \end{subfigure}
    \caption{A cross section of a DPTS pixel and a photo of the chip under a microscope.
    }
    \label{fig:dpts:overview}
\end{figure}

The Digital Pixel Test Structure (DPTS) is the most complex prototype MAPS produced in the first submission in TPSCo \SI{65}{\nm} technology process~\cite{tower}.
In order to optimise the technology for ionising particle detection, the submission was produced in four process splits, gradually modifying the doping levels of various implants (cf.~Fig.~\ref{fig:dpts:cross_section}). The present work focuses only on the split expected to yield the best performance \cite{snoeys-pos}.

To better collect the signal charge and accelerate it to the collection diode, similar measures to those in the \qty{180}{\nm} technology have been applied \cite{malta2,modproc}.
As illustrated in Fig.~\ref{fig:dpts:cross_section}, a deep, low-dose, \textit{n}-type implant has been introduced to displace the junction from the collection diode into the epitaxial layer and thus deplete the epitaxial layer over the full pixel width \cite{modproc}.
This additional implant does not extend to the pixel border, but there is a gap in the implant near the pixel edges with the aim to increase the lateral field, pushing the signal charge to the collection diode \cite{munker}.
This is expected not only to accelerate the charge collection but also to reduce charge sharing to give more operating margin due to the larger seed pixel signal. 

\begin{figure}[!ht]
  \centering
  \includegraphics[width=0.85\textwidth=false]{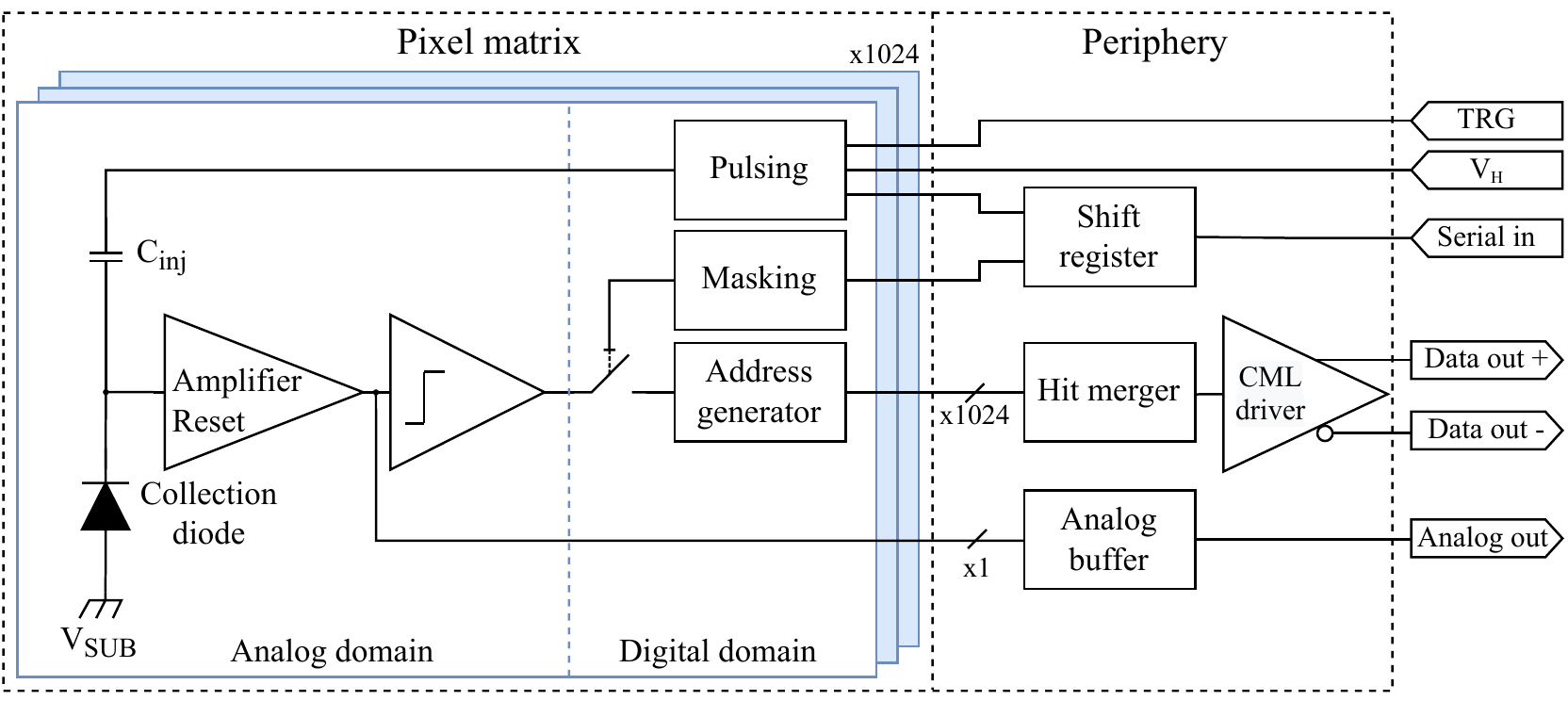}
  \caption{DPTS functional diagram. \num{1024} pixels can be masked from readout and selected for pulsing via a shift register. The addresses of hit pixels are read out via a differential digital output line. The  in-pixel amplifier output of a single pixel is connected to an interface pad.}
  \label{fig:dpts:chip_scheme}
\end{figure}

The DPTS chip (cf.~Fig.~\ref{fig:dpts:photo}) measures \qtyproduct{1.5x1.5}{\mm} and features a \numproduct{32x32} pixel matrix of \qtyproduct{15x15}{\um} pitch, controlled by a set of external reference currents and voltages, and read out via a current mode logic (CML) output. 
The in-pixel front end amplifies, shapes, and discriminates the signal from the collection diode. The positions of the hit pixels are time-encoded (c.f.~Sec.~\ref{sec:dpts:encoding}) and all 1024 pixels are read out simultaneously via a differential digital output line (cf.~Fig.~\ref{fig:dpts:chip_scheme}).
The pixels can be masked from readout and selected for pulsing via a \qty{480}{\bit} triplicated shift register. Given its space-constrained size, the possible masking and pulsing combinations are limited. Those used in this paper are the following: pulsing a single pixel, masking a single pixel, and masking the whole matrix with the exception of one row (unmasking a single pixel is not possible).

Figure~\ref{fig:dpts:photo} shows a photograph of the DPTS taken under a microscope. Starting from the centre and going outwards, the pixel matrix, shift register block, guard rings, and bonding pads can be distinguished. Most of the pads are dedicated to supplying power and reverse bias to the chip. The power supply is split into three domains (all operated at \qty{+1.2}{\V}): analogue in-pixel front end, CML output driver, and all other digital circuitry. The reverse bias can be supplied separately to the substrate ($\Vsub$) and the deep p-wells ($\Vpwell$) hosting the circuitry (cf.~Fig.~\ref{fig:dpts:cross_section}), however, in this work they were kept at the same potential. 

There are three variants of the chip, implementing different time-encoding (cf.~Sec.~\ref{sec:dpts:encoding}) and ground connection schemes\footnote{In one of the DPTS variants, the ground of the digital and analogue power domains is connected on the chip. In all the measurements presented in this paper, the ground was connected for all three power domains off-chip.}. As the laboratory testing revealed no difference in their performance, no distinction will be made in the results discussed here.

\subsection{In-pixel front end}
\label{sec:dpts:frontend}

\begin{figure}[!ht]
  \centering
  \includegraphics[width=0.9\textwidth=false]{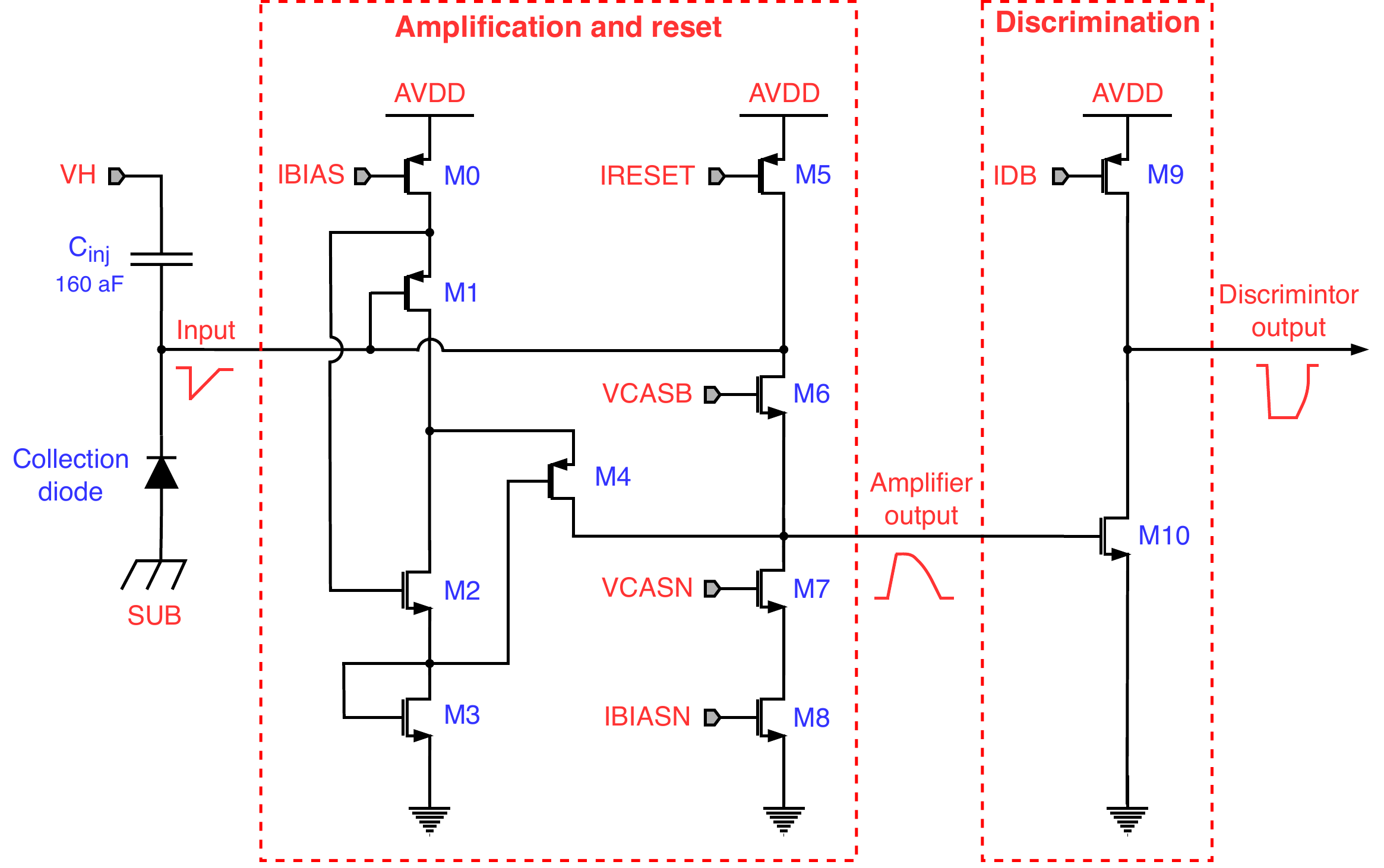}
  \caption{In-pixel amplifying, discriminating and reset circuit. All front-end biases are provided externally, in common to all pixels.}
  \label{fig:dpts:pixel_scheme}
\end{figure}

The analogue in-pixel front-end circuit (cf.~Fig.~\ref{fig:dpts:pixel_scheme}) is controlled via four currents ($\Ib$, $\Ibn$, $\Ir$ and $\Id$) and two voltages ($\Vb$ and $\Vn$), all externally generated and supplied to the chip via the interface pads.
The front end \cite{piro-nss} is based on a high gain cascoded inverting amplifier, requiring direct feedback to the input to correctly define its operating point. This is achieved by the $\Vb$ transistor (\texttt{M6}) in combination with the $\Ir$ current source (\texttt{M5}): the feedback will make the current through the $\Vb$ transistor equal to $\Ir$ minus the leakage current on the collection electrode and correctly set the voltage of the collection electrode.
When charge is collected on the collection diode, the output of the inverting amplifier at the source of the $\Vb$ transistor will make a positive voltage excursion and switch off the $\Vb$ transistor, causing the collection diode to be reset by a constant current and, therefore, resulting in a close to linear time-over-threshold behaviour.
The high gain amplifier output is the input to a subsequent common-source stage (\texttt{M9-M10}). As soon as the amplifier output is sufficiently high for the input transistor of this second stage to overcome $\Id$, the output node of the second branch will fall and convert the signal into a digital rail-to-rail signal.

The front end has been designed to operate at current levels in the main branch ($\Ib$, \texttt{M0-M3}) between \qty{10}{\nA} and \qty{5}{\uA}, in a power-to-speed trade-off. The $\Ibn$ current source (\texttt{M8}) balances the amplifier output current and is to be operated at one-tenth of the $\Ib$ current. $\Vn$ (\texttt{M7}) is used to further control the operating margin of the amplifier.
In the lower current ranges, all transistors are in weak inversion. In addition, the NMOS transistors are subject to the reverse bias applied to the sensor ($\Vpwell$), requiring biases like $\Vb$ and $\Vn$ to be adjusted to $\Vpwell$ voltage and current levels in the circuit.
The power consumption in the steady state is given by the current in the main branch ($\Ib$) as the reset current is orders of magnitude lower and the $\Id$ current in the discriminator branch is flowing only while \texttt{M10} is active. With the typical settings used in this paper (c.f.~Sec.~\ref{sec:daq_setup}), it amounts to about \qty{120}{\nano\watt}.

A test circuitry, which can inject charge in the collection electrode through a capacitance of $C_p=\qty{160}{\atto\F}$, is also integrated in the pixel (cf.~Fig.~\ref{fig:dpts:chip_scheme} and Fig.~\ref{fig:dpts:pixel_scheme}). The amount of injected charge can be regulated by an external voltage reference, $\Vh$, and the injection is triggered by asserting the TRG signal via an interface pad.

Besides the 1024 digital output pixels, the DPTS features a monitoring pixel with an analogue output connected to an interface pad. The analogue pixel front end is identical to the one described above with the exception of the last common-source stage being replaced by a source follower, which buffers its analogue output signal to the interface pad.

\subsection{Hit position encoding and time-over-threshold}
\label{sec:dpts:encoding}

\begin{figure}[!ht]
  \centering
  \includegraphics[width=0.99\textwidth]{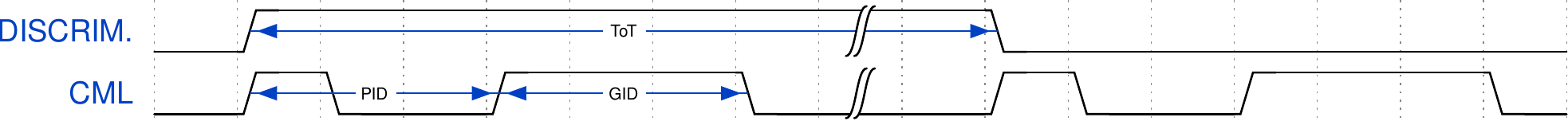}
  \caption{DPTS hit position encoding scheme. The two pulses encoding the pixel position with the matrix are sent to CML output, both at the assertion and at the deassertion of the discriminated pixel output signal.}
  \label{fig:encoding}
\end{figure}

The transition of the discriminator output triggers the address generator (cf.~Fig.~\ref{fig:dpts:chip_scheme}) to send two consecutive pulses on the CML output with a duration based on the pixel position \cite{cecconi-twepp}. As indicated in Fig.~\ref{fig:encoding}, the first pulse is of a fixed duration, the time distance between the two pulses encodes the pixel position in a group of columns (PID) and the duration of the second pulse the column group position in the matrix (GID). 

The assertion and the deassertion of the discriminator result in two sets of pulses, delimiting the time interval in which the front-end pulse was over the threshold (ToT). Because the front end is designed such that the pulse length is monotonically increasing with the input signal, the ToT provides information on the collected or injected charge.

As the time-encoded output of the 1024 pixels is merged to a single output line, two or more pixels firing simultaneously result in a signal collision, i.e. non-decodable output (the firing pixels' coordinates cannot be determined).
In order to minimise the probability of signal collisions in case of charge sharing among neighbouring pixels, two optimisations were implemented in the encoding scheme: (1) the output signals of pixels in every other column are delayed by a fixed offset, and (2) in every other row of a column pair, the GID is swapped between the two pixels.
\section{Laboratory measurements}
\subsection{Data acquisition setup}
\label{sec:daq_setup}

The laboratory and testbeam measurements used a custom-designed setup for the DPTS sensor that supplies biases and control signals to the chip. The CML and analogue outputs of the chip were recorded on an oscilloscope, with a sampling rate of 5~GS/s and a bandwidth of \qty{500}{\MHz}. 
For all the temperature sensitive measurements, in particular those involving irradiated chips, the chip temperature was controlled using a water-cooled (+\SI{20}{\celsius}) aluminium jig in thermal contact with the chip carrier card and thus the chip.

The bias settings of the chip are set to ``nominal'' values (defined by the design operating point) for all measurements unless otherwise stated or varied in the measurement. These settings are $\Vb = \qty{300}{\mV}$, $\Vn = \qty{300}{\mV}$, $\Ir = \qty{10}{\pA}$, $\Id = \qty{100}{\nA}$, $\Ib = \qty{100}{\nA}$, $\Ibn = \qty{10}{\nA}$, $\Vh = \qty{600}{\mV}$ (the pixel pulsing amplitude) and $\Vsub = \Vpwell = \qty{-1.2}{\V}$. For irradiated sensors, the $\Ir$ is increased to \SI{35}{\pA} to overcome the sensor leakage current. Furthermore, when adjusting $\Vpwell$, the values of $\Vb$ and $\Vn$ are adjusted as well (cf.~Sec.~\ref{sec:dpts:frontend}).
The maximum $\Vpwell$ foreseen by the design is $\qty{-6}{\volt}$, however, in this work, the $\Vpwell$ is capped at $\qty{-3}{\volt}$ as further increasing it did not improve performance, at least not in the measurements presented here.

\subsection{Analogue response}
\label{sec:lab_meas_analogue}
\begin{figure}[!ht]
    \centering
    \includegraphics[width=0.95\textwidth]{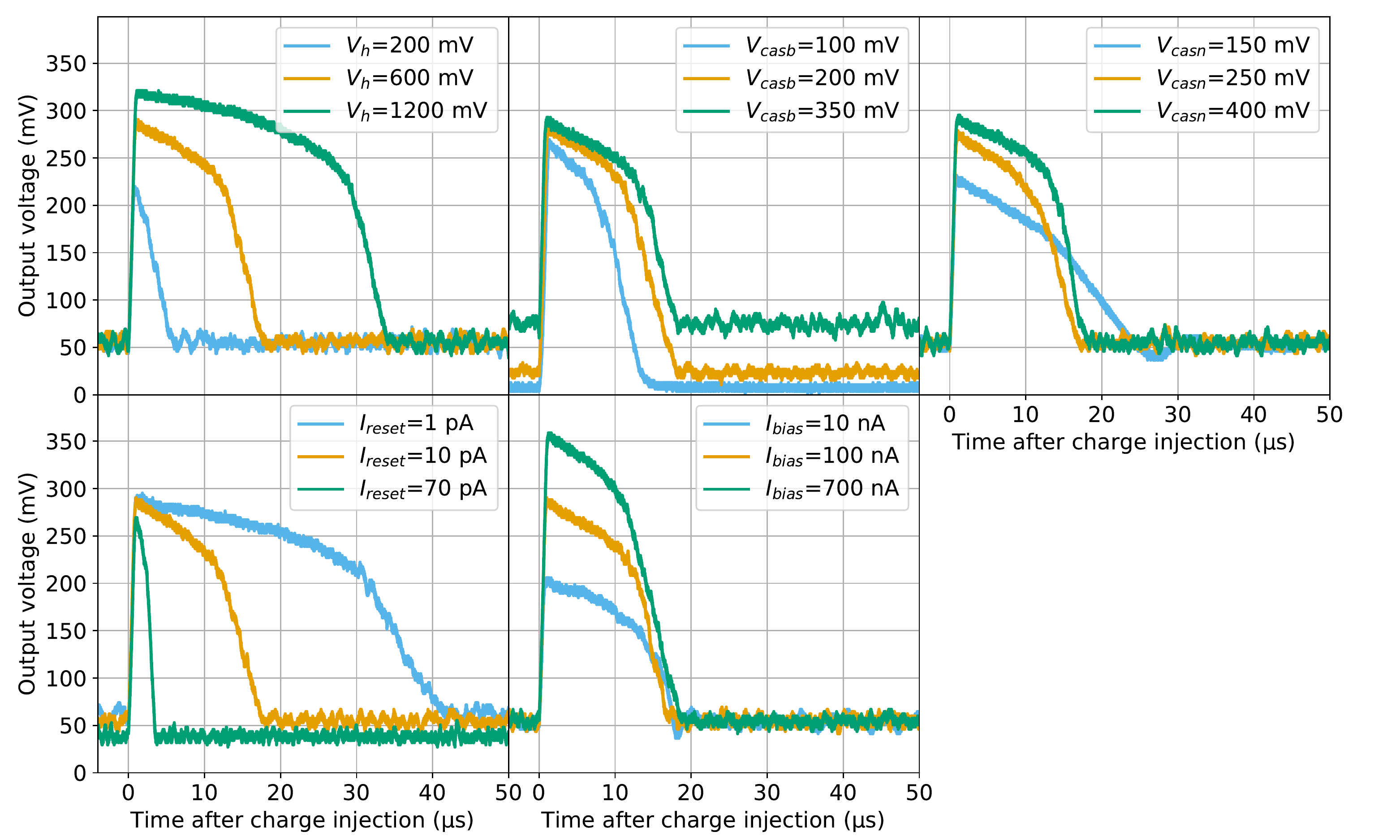}
    \caption{The analogue responses produced when the monitor pixel is pulsed and $\Vh$, $\Vb$, $\Vn$, $\Ir$ and $\Ib$ are varied. For $\Ib$, $\Ibn = \Ib/10$. Other chip bias settings not mentioned in the legend are at nominal values (cf.~Sec.~\ref{sec:daq_setup}).}
    \label{fig:aout}
\end{figure}

The inclusion of an additional monitor pixel to the chip allows the analogue pulse of the front end to be directly measured (cf. Sec. \ref{sec:dpts:frontend}).
The monitor pixel was pulsed, and the output responses at different chip bias settings are shown in Fig.~\ref{fig:aout}, demonstrating the influence of varying $\Vh$, $\Vb$, $\Vn$, $\Ir$, $\Ib$ and $\Ibn$. For the case of $\Ib$, $\Ibn$ was also varied, keeping $\Ibn = \Ib/10$, which is the recommended regime

Increasing $\Vh$, which controls the injected charge, increases the amplitude and length of the pulse. It can easily be observed that, in accordance with the amplifier design (c.f.~Sec.~\ref{sec:dpts:frontend}) the pulse length, rather than amplitude, is proportional to the amount of injected charge.
$\Vb$ has a direct influence on the baseline, at low $\Vn$ the pulse changes shape, $\Ir$ changes the duration of the pulse and increasing $\Ib$ increases the amplitude. The impact of changing $\Id$ on the analogue pulse is not presented as this last stage of the front end is not included in the monitor pixel.

\subsection{Threshold and noise}
\label{sec:lab_meas_thr}
\begin{figure}[!htb]
    \centering
    \includegraphics[width=0.79\textwidth]{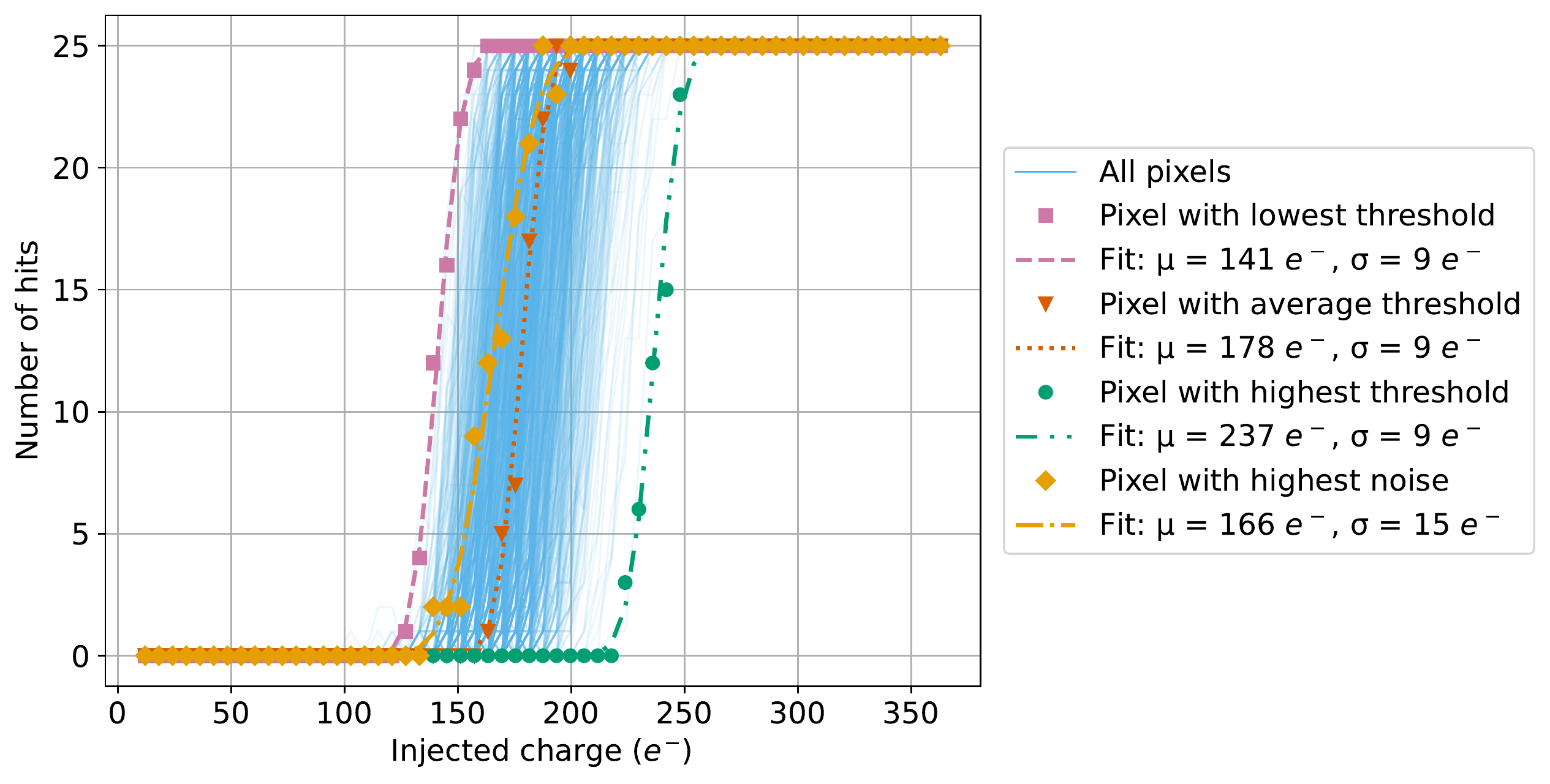}
    \caption{S-curves of a chip tuned to a threshold of \SI{180}{\ele} with $\Vb=\qty{180}{\mV}$ at $\Vsub=\qty{-1.2}{\V}$. Four pixels have been highlighted along with their s-curve fits (non-solid lines). These four pixels represent the response of pixels with the lowest, average and highest threshold and the pixel with the highest noise. Other chip bias settings are at nominal values (cf.~Sec.~\ref{sec:daq_setup}).}
    \label{fig:s-curve}
\end{figure}

Investigation into the front-end response of the chip used the in-pixel pulsing circuitry to conduct threshold measurements. This involved counting the number of hits while varying $\Vh$, that is, the amount of injected charge. For every $\Vh$ value, each pixel was pulsed 25~times. As the injected charge approaches the threshold, the number of hits increases until a plateau is reached, producing a so-called ``s-curve''. Figure \ref{fig:s-curve} shows examples of the measured s-curves. The threshold and the noise are given by the mean and standard deviation of the derivative of the s-curve, respectively. A hit was determined to occur if at least two sets of pulses were captured by the oscilloscope, ensuring both the assertion and deassertion of the discriminator were recorded (cf.~Sec.~\ref{sec:dpts:encoding}). The final threshold value is calibrated to electrons via a measurement of the $\text{Mn-K}_{\alpha}$ peak position, details of which are given in Sec.~\ref{sec:lab_meas_55Fe}, and this method is used throughout the paper. This calibration is done separately for each sensor, as deviations of up to \qty{20}{\percent} of the measured injection capacitance with respect to its design value have been observed.

\begin{figure}[!ht]
    \centering
    \includegraphics[width=0.99\textwidth]{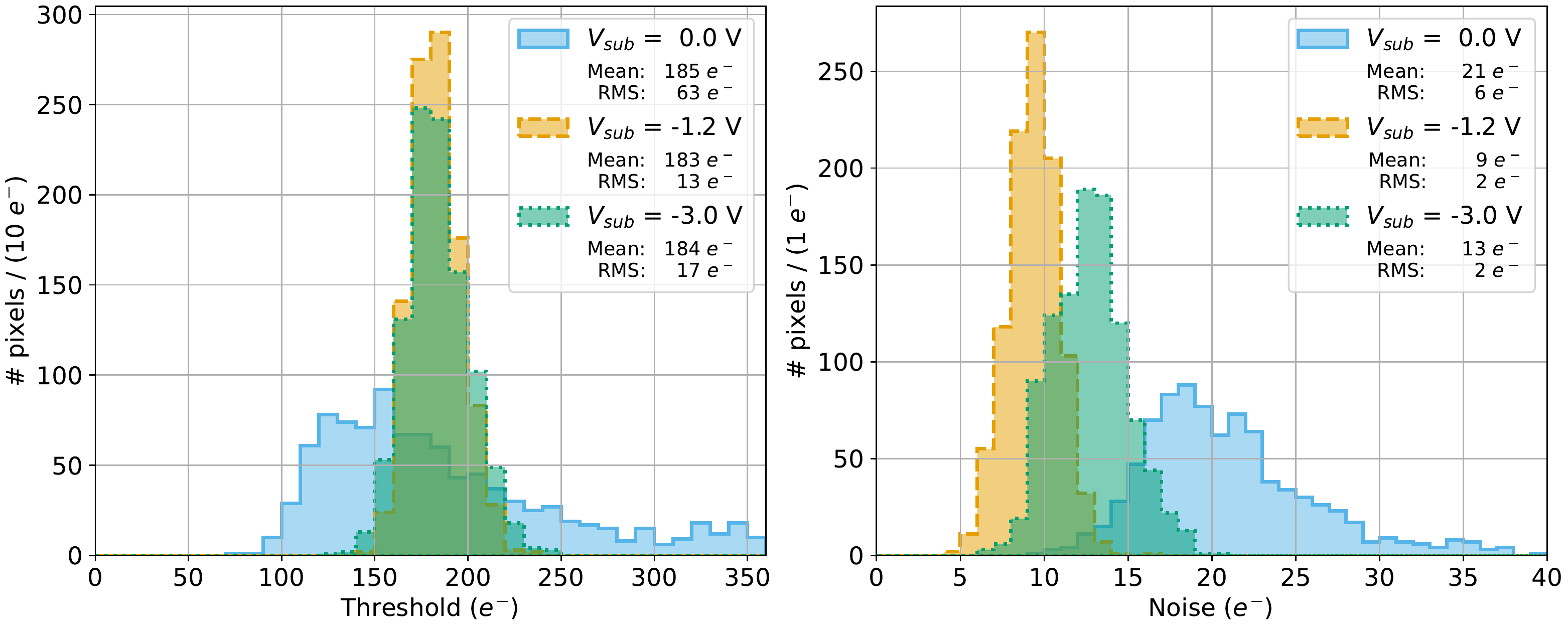}
    \caption{Pixel threshold (left) and noise (right) distributions of a chip tuned to a threshold of \SI{180}{\ele} with $\Vb=\qtylist[list-units=single]{110;180;500}{\mV}$ at $\Vsub = \qtylist[list-units=single]{0.0;-1.2;-3}{\V}$, respectively. Other chip bias settings are at nominal values (cf.~Sec.~\ref{sec:daq_setup}). }
    \label{fig:thr_histo}
\end{figure}

Examples of the pixel threshold and noise distributions at $\Vsub = \qty{0}{\V}$, \SI{-1.2}{\V} and \SI{-3}{\V} are shown in Fig.~\ref{fig:thr_histo}, with the $\Vb$ value chosen such that the mean threshold of all pixels is \SI{180}{\ele}. The threshold distributions show comparable performance for $\Vsub = \qtylist{-1.2;-3}{\V}$, with both having a similar width and centred around \SI{180}{\ele}. However, at $\Vsub = \qty{0}{\V}$, the distribution is wider and asymmetric -- indicating a substantial pixel-to-pixel threshold spread resulting in a non-uniform response across the matrix. For the noise distributions, $\Vsub = \qty{0}{\V}$ continues to show the widest distribution, while $\Vsub = \qty{-1.2}{\V}$ has a better performance in terms of a smaller mean and spread than $\Vsub = \qty{-3}{\V}$, suggesting non-optimal chip biasing settings for $\Vsub = \qty{-3}{\V}$. In general, even with chip biasing settings optimised for threshold and noise, a worse performance in terms of threshold spread and noise was observed when decreasing $\Vsub$ towards \SI{0}{\V} or increasing it towards \SI{-3}{\V} and above.

\begin{figure}[!ht]
    \centering
    \includegraphics[width=0.99\textwidth]{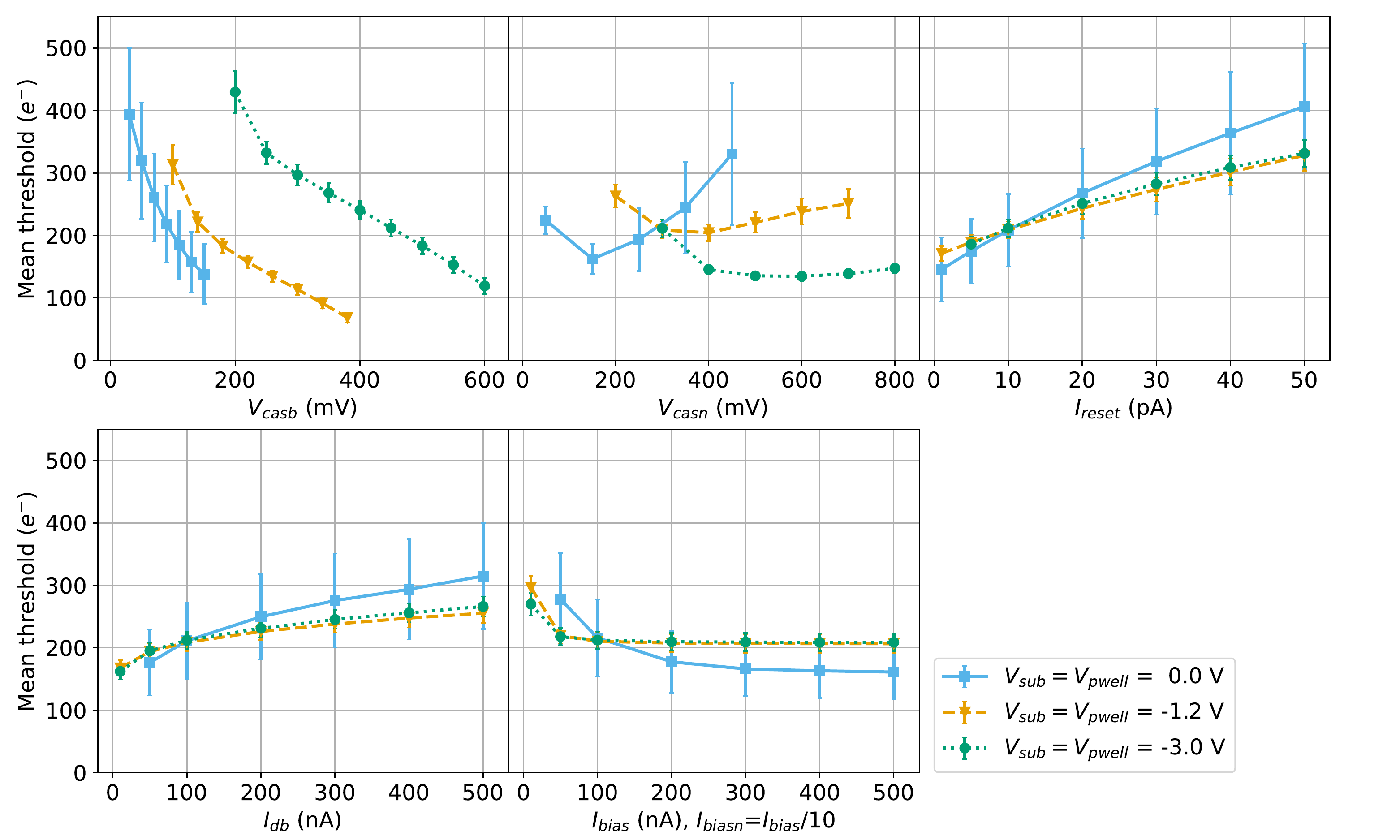}
    \caption{The mean threshold as a function of different chip bias settings at $\Vsub$ of $\qtylist[list-units=single]{0.0;-1.2;-3}{\V}$ with $\Vb$ set to \qtylist[list-units=single]{95;150;450}{\mV}, respectively. $\Vb$ was tuned so that the mean threshold was \SI{210}{\ele} for nominal settings. The error bars represent the threshold RMS. The parameters not varied in the measurement are at nominal values (cf.~Sec.~\ref{sec:daq_setup}).}
    \label{fig:thr_mean}
\end{figure}

Figure~\ref{fig:thr_mean} demonstrates the effect of the chip bias settings on the threshold at $\Vsub = \qtylist{0;-1.2;-3}{\V}$ by plotting the mean and the RMS (given by the error bars) of all pixels. For $\Vb$, it can be observed that this chip bias has the largest impact on the threshold. Also, there is a linear relationship between the mean threshold and $\Vb$ in a certain range, which is larger for larger $\Vsub$. This linear relationship is the motivation for using $\Vb$ as the main handle to control the threshold of the chip during operation and for all subsequent results. Figure~\ref{fig:thr_mean} also illustrates that at different $\Vsub$ values, the same value of $\Vb$ does not produce the same mean threshold. 

While the other chip biases affect the threshold as well, for $\Vn$, $\Ib$ and $\Ibn$, the influence is expected to be minimal once in a stable operating regime. The changes due to chip biases other than $\Vb$ are largest at $\Vsub = \qty{0}{\V}$. Whereas, at lower reverse bias values, the changes to the mean threshold (for the plotted range) are within \qty{100}{\ele} except for $\Ir$, where the increase is around \qty{150}{\ele}. However, this large dependence of the threshold on $\Ir$ does not mean that this parameter is a good handle to set the threshold because changing $\Ir$ alters the diode biasing (cf.~Sec.~\ref{sec:dpts:frontend}) and influences the charge collection properties. Consequently, $\Ir$ is only used to compensate for the increased leakage current caused by non-ionising radiation damage and not to tune the threshold.

Another operating condition that impacts the threshold is the temperature. Measurements of a non-irradiated chip have shown the mean threshold decreases by \SI{0.5}{\ele} per degree Celsius in the measured range of \SIrange{15}{40}{\celsius}. However, for an irradiated chip, the dependence is not linear. This is compatible with the non-linear dependence of the leakage current on the temperature for higher non-ionising radiation fluence levels.

\subsection{Time-over-threshold}
\label{sec:lab_meas_tot}

\begin{figure}[!ht]
    \centering
    \includegraphics[width=0.55\textwidth]{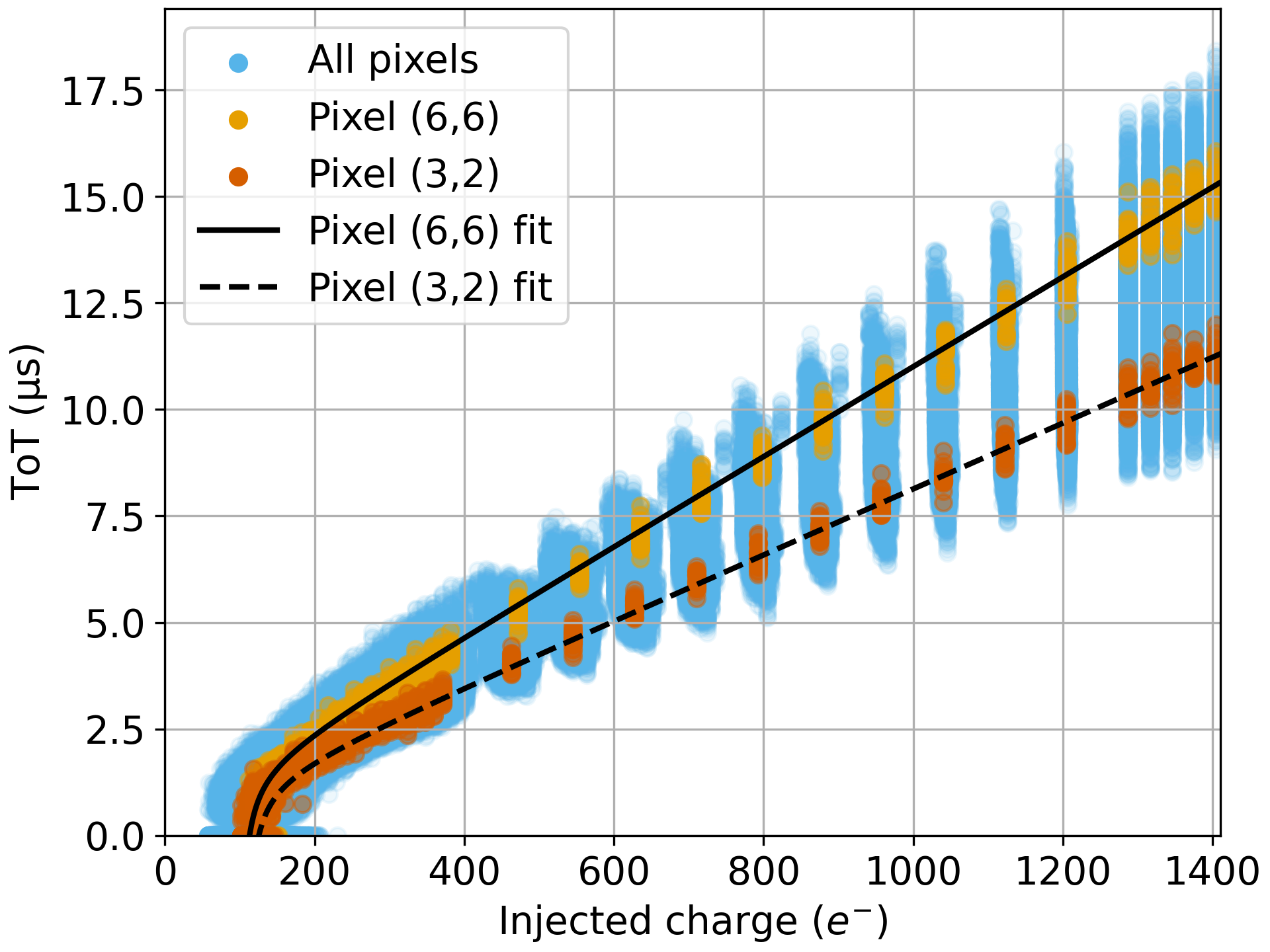}
    \caption{Time-over-threshold (ToT) as a function of injected charge for all pixels with two pixels highlighted to demonstrate the pixel-to-pixel variation. The two pixels have been fitted with Eq.~\ref{eq:ToT_vs_vh} shown by the solid and dashed lines.
    }
    \label{fig:ToT_vs_VH}
\end{figure}

Further investigation into the front-end response was performed by observing the impact of the chip biasing settings on the time-over-threshold (ToT, cf.~Sec.~\ref{sec:dpts:encoding}). By varying the injected charge, $Q_{inj}$, and measuring the ToT for all pixels, it can be seen from Fig.~\ref{fig:ToT_vs_VH} that at lower injected charge values, around the pixel threshold, the response is non-linear. Whereas, for larger injected charge values, the ToT response becomes linear and the spread in the ToT increases. There is also a sizeable pixel-to-pixel variation that is demonstrated by highlighting the response of two pixels. To account for this spread, a ToT calibration was performed for all pixels individually, by fitting with the empirical function

\begin{equation}
    \label{eq:ToT_vs_vh}
    ToT = aQ_{inj} + b - \frac{c}{Q_{inj}-d}\text{ ,}
\end{equation}

\noindent where $a$, $b$, $c$ and $d$ are the fit parameters. Example fits of two pixels are shown by the solid lines in Figure~\ref{fig:ToT_vs_VH}. The effect of the ToT calibration to normalise the pixel response over the matrix is shown and discussed in Sec.~\ref{sec:lab_meas_55Fe}.

\begin{figure}[!hb]
    \centering
    \includegraphics[width=0.99\textwidth]{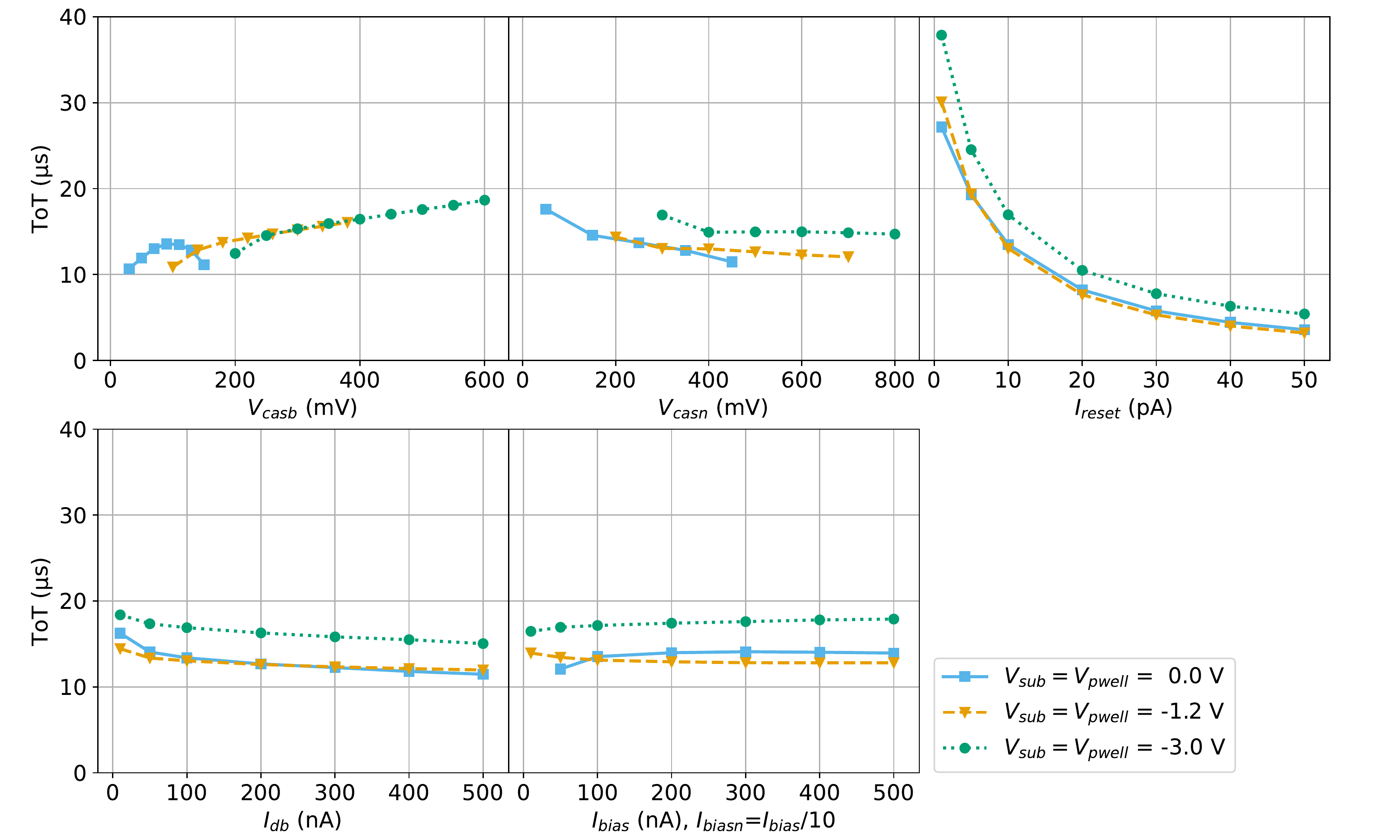}
    \caption{Time-over-threshold (ToT) as a function of different chip biases at a fixed charged injection of \SI{725}{\ele}. Non-varied parameters are at nominal values (cf.~Sec.~\ref{sec:daq_setup}).}
    \label{fig:ToT}
\end{figure}

The impact of the other chip biases on ToT was also investigated at a fixed charged injection of \SI{725}{\ele}. The results in Fig.~\ref{fig:ToT} demonstrate that $\Ir$ produces the most significant change in ToT (within \SI{40}{\micro\s}), which is compatible with $\Ir$ resetting the diode (cf.~Sec.~\ref{sec:dpts:frontend}). For the other biases, the change is within \SI{10}{\micro\s}. Furthermore, the non-linear behaviour of the ToT as a function of $\Vb$ for values above \SI{70}{\mV} illustrates the limited operation margin at $\Vsub = \qty{0}{\V}$.

\subsection{Fake-hit rate}
\label{sec:lab_meas_fhr}

\begin{figure}[!hb]
    \centering
    \includegraphics[width=0.99\textwidth]{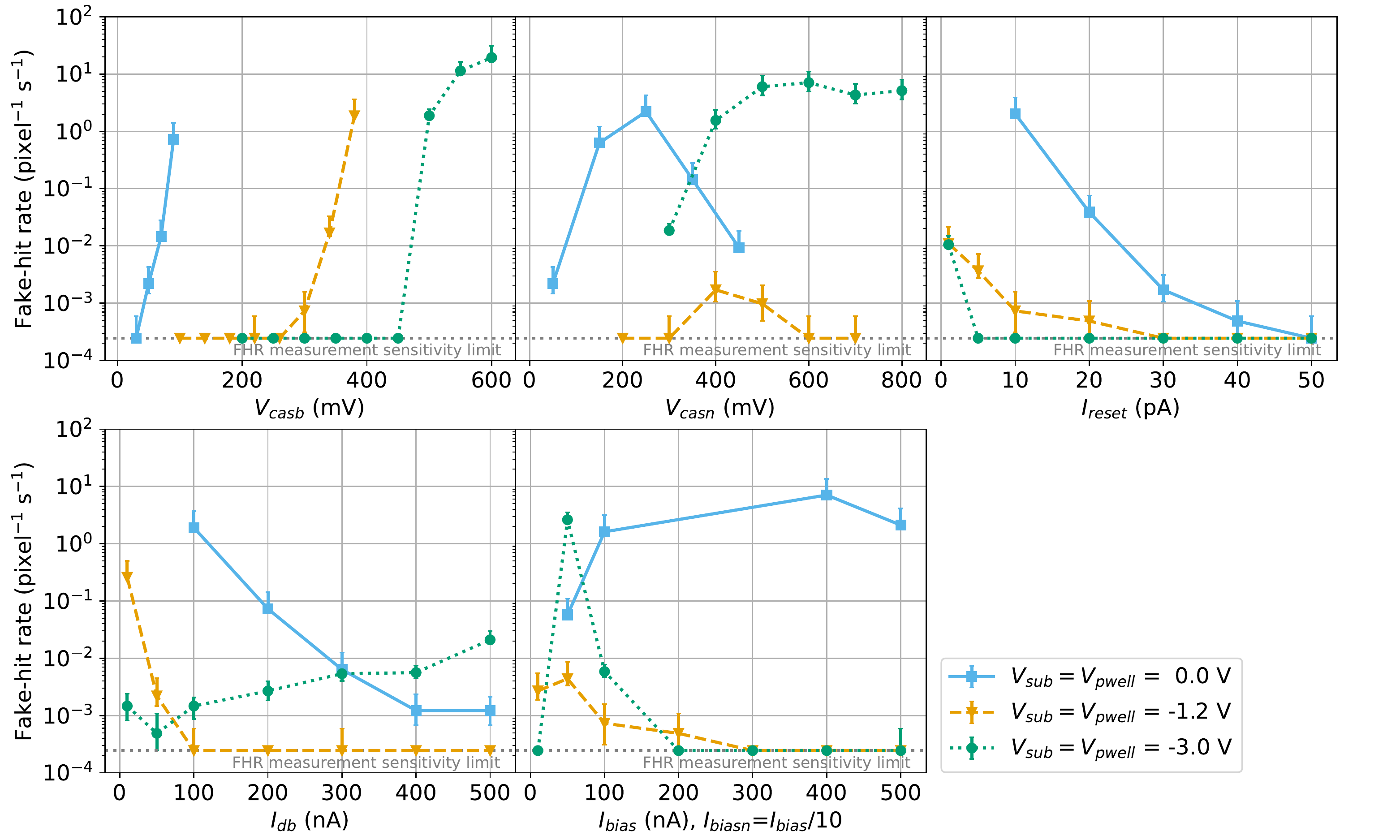}
    \caption{Fake-hit rate as a function of  different chip biases. At $\Vsub=\qty{-1.2}{\V}$, the threshold was set at \SI{120}{\ele}, while for the other $\Vsub$ values it was \SI{210}{\ele} (for the nominal parameters). Other chip bias settings not varied in the measurement are at nominal values (cf.~Sec.~\ref{sec:daq_setup}).}
    \label{fig:FHR}
\end{figure}

The impact of the chip bias settings on the fake-hit rate (FHR) was evaluated and is shown in Fig.~\ref{fig:FHR}. The FHR is defined as the number of hits per pixel and second in the absence of external stimuli as the chip operates in a continuous readout.
Given that the oscilloscope captures a fixed-size time window, the FHR is estimated as the aggregate of hits in a number of randomly triggered oscilloscope acquisitions divided by the sum of their duration and the total number of pixels (\num{1024}).
In general, the measurements show that the lower the threshold, the higher the FHR. As with the threshold results shown in Sec.~\ref{sec:lab_meas_thr}, operating at $\Vsub=\qty{-1.2}{\V}$ gives the best performance, with many of the results below the sensitivity limit\footnote{The sensitivity limit is defined as the lowest measurable FHR for the number of oscilloscope captures in the measurement.} of the measurements, even at the low threshold used for the measurements. At $\Vsub=\qty{0}{\V}$, the sharp increase in the FHR from the lowest operable $\Vb$ further illustrates the limited operation margin without supplying reverse bias.

\subsection{Pixel position decoding}
\label{sec:lab_meas_decode}

\begin{figure}[!ht]
    \centering
    \includegraphics[width=0.6\textwidth]{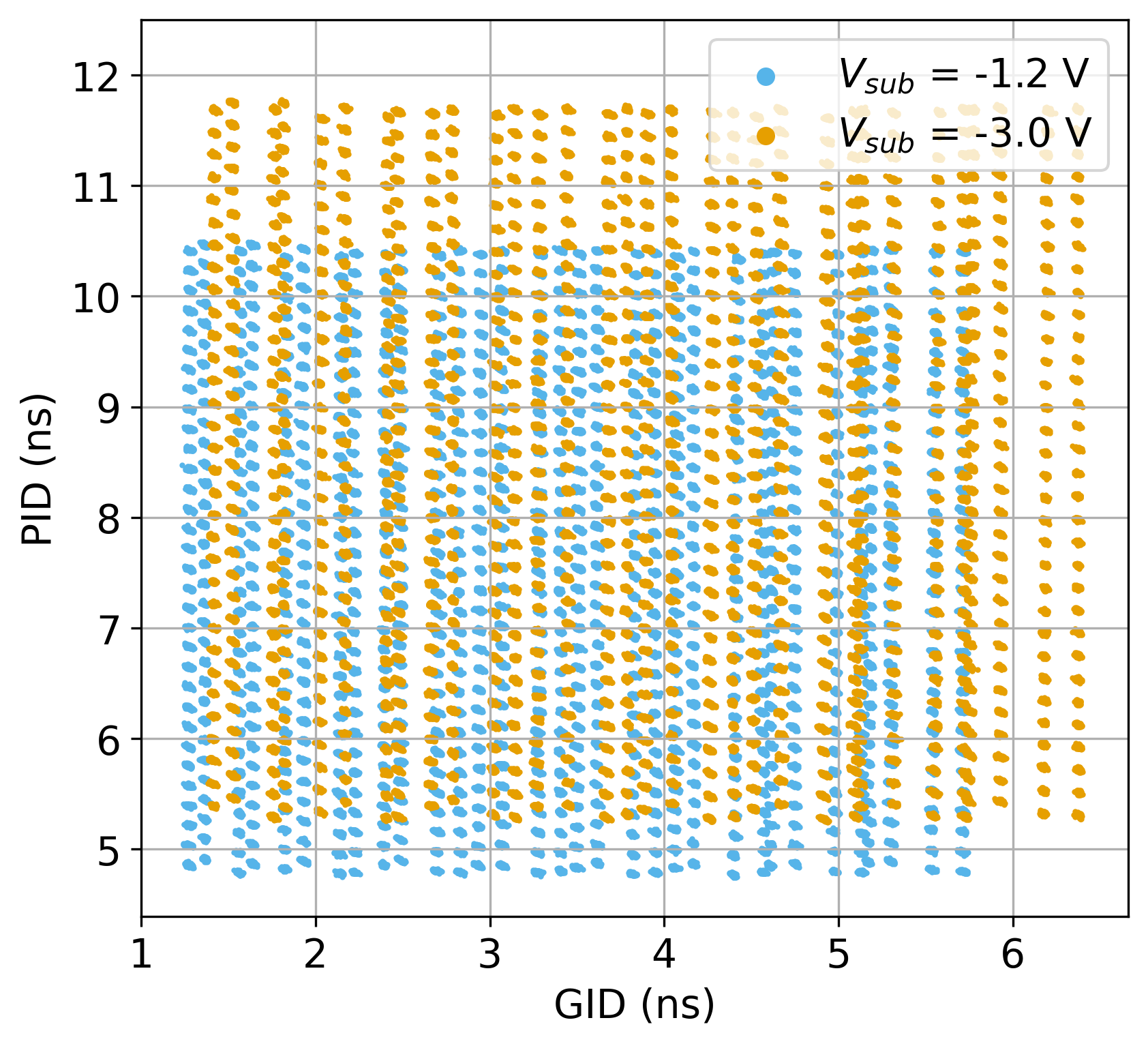}
    \caption{The PID and GID data points corresponding to all pixels pulsed 100 times at $\Vsub=\Vpwell=\qtylist{-1.2;-3}{\V}$ demonstrating the reverse bias dependence. Each cluster consists of 100 points and represents the measured PID and GID value for one pixel. 
    }
    \label{fig:decoding}
\end{figure}

As described in Sec.~\ref{sec:dpts:encoding}, the positions of the hit pixels are time-encoded in the CML output pulses in terms of PIDs and GIDs. An example of the measured PIDs and GIDs for all pixels, each pulsed 100 times, at $\Vsub = \Vpwell = \qtylist{-1.2;-3}{\V}$ is given in Fig.~\ref{fig:decoding}.
It can be observed that the data points form clusters, corresponding to 1024 pixels.
Some of the clusters for the different pixels are very close (and look to be overlapping due to an artefact of the axis scales), demonstrating where the decoding errors can arise.

Given that the time constants in the encoding are dependent on the production process and measurement conditions \cite{cecconi-twepp}, a decoding calibration is needed to convert the measured PIDs and GIDs of the waveforms to pixel rows and columns. The calibration is performed by pulsing each pixel 100 times and measuring the PIDs and GIDs of the recorded waveforms. The centre of gravity of the PID and GID cluster for each pixel gives the decoding calibration. By associating a (GID, PID) pair to the nearest centre of gravity, the conversion to columns and rows is obtained.

Except for the reverse bias dependence, PIDs and GIDs were found to also depend on supply voltage and temperature. 
In particular, it was found that the mean PID and GID increase by \qtylist{8;4}{\ps} per degree Celsius, respectively, in the range \SIrange{15}{40}{\celsius}.

\subsection{\texorpdfstring{Measurements with X-rays emitted by an \textsuperscript{55}Fe source}{Measurements with X-rays emitted by an 55Fe source}}
\label{sec:lab_meas_55Fe}

\begin{figure}[!hbt]
    \centering
    \includegraphics[width=0.75\textwidth]{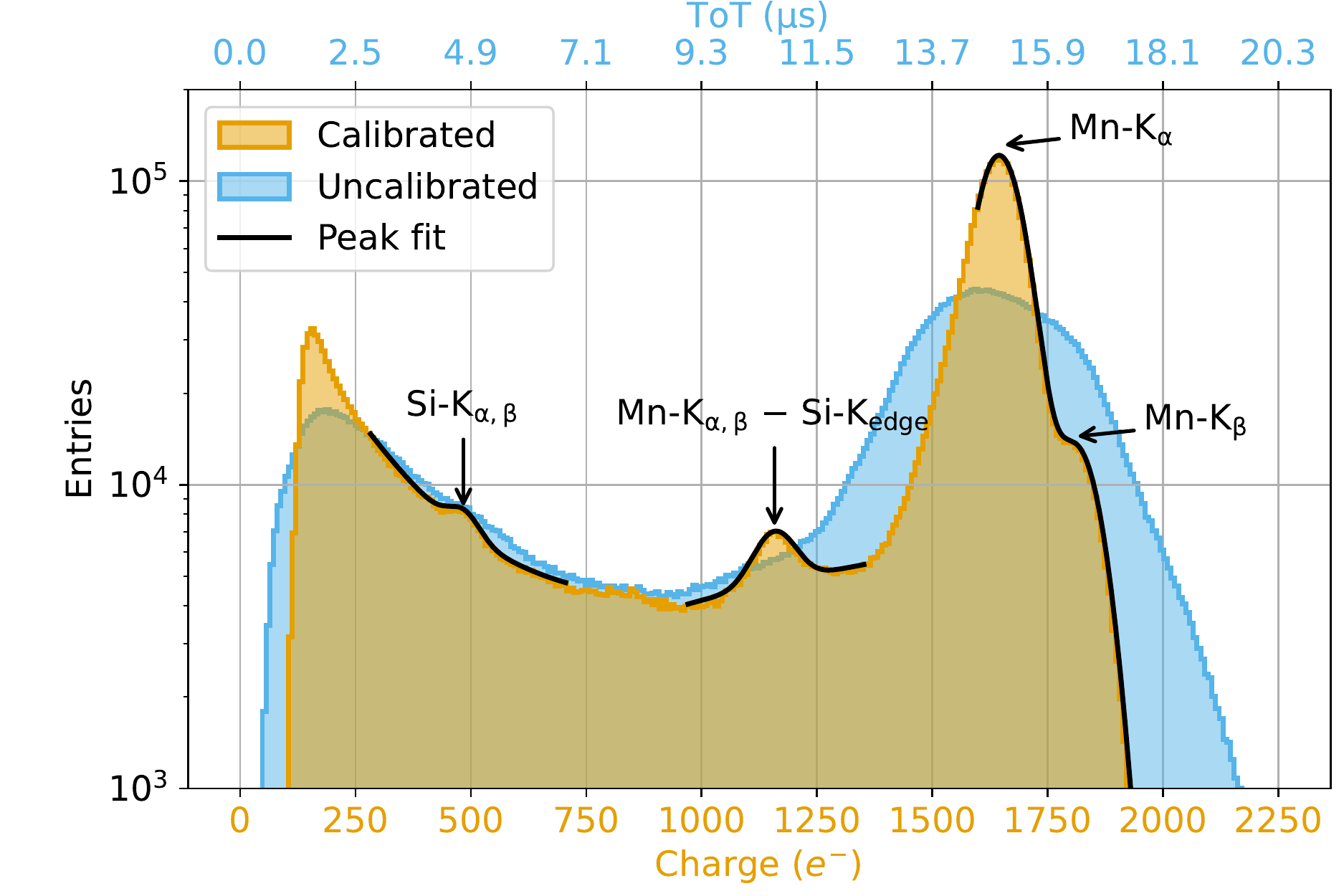}
    \caption{Measured $^{55}$Fe spectrum of single pixel clusters with a threshold set to \SI{120}{\ele}. The initial spectrum (blue) is ToT calibrated (orange) which resolves the two x-ray peaks ($\text{Mn-K}_{\alpha}$ and $\text{Mn-K}_{\beta}$) plus the $\text{Mn-K}_{\alpha,\beta}$ escape ($\text{Mn-K}_{\alpha,\beta} - \text{Si-K}_{edge}$) and silicon fluorescence ($\text{Si-K}_{\alpha,\beta}$) peaks.}
    \label{fig:fe55}
\end{figure}

X-ray emissions from an $^{55}$Fe source illuminating the top side of the chip at a distance of approximately \SI{12}{\mm} have been measured. Two cuts have been applied to the data: the removal of all pixels on the matrix edge and the removal of pixels with hit rates of five standard deviations above the mean. The measured signal spectrum of clusters involving only one pixel for a sensor with a threshold tuned to \SI{120}{\ele} is shown in Fig.~\ref{fig:fe55} (in blue).

A ToT calibration for each pixel is applied to account for the variation in the pixel-to-pixel response, as described in Sec.~\ref{sec:lab_meas_tot}. After the ToT calibration, the $\text{Mn-K}_{\alpha}$ and $\text{Mn-K}_{\beta}$ emission peaks are resolved, as shown in Fig.~\ref{fig:fe55} (in orange), as well as the $\text{Mn-K}_{\alpha,\beta}$ escape ($\text{Mn-K}_{\alpha,\beta} - \text{Si-K}_{edge}$) and silicon fluorescence ($\text{Si-K}_{\alpha,\beta}$) peaks. The $\text{Mn-K}_{\alpha}$ and $\text{Mn-K}_{\beta}$ peaks are fitted with a sum of two Gaussians. The resolution of the $\text{Mn-K}_{\alpha}$ peak is defined as the FWHM divided by the measured peak value and was calculated to be $\qty{7.42}{\percent}\pm\qty{0.01}{\percent}$. For the fluorescence and escape peaks, the signal is fitted with a Gaussian and a background fit of either exponential or linear form for the fluorescence and escape peaks, respectively.
Since the response of ToT as a function of injected charge was found to be linear for ToT values above $\qty{3}{\us}$ (cf.~Sec.~\ref{sec:lab_meas_tot}), the comparison between the measured and literature peak values is fitted with a linear function, giving the conversion from ToT to energy in \SI{}{\ele}.

\begin{figure}[!hbt]
    \centering
    \includegraphics[width=0.75\textwidth]{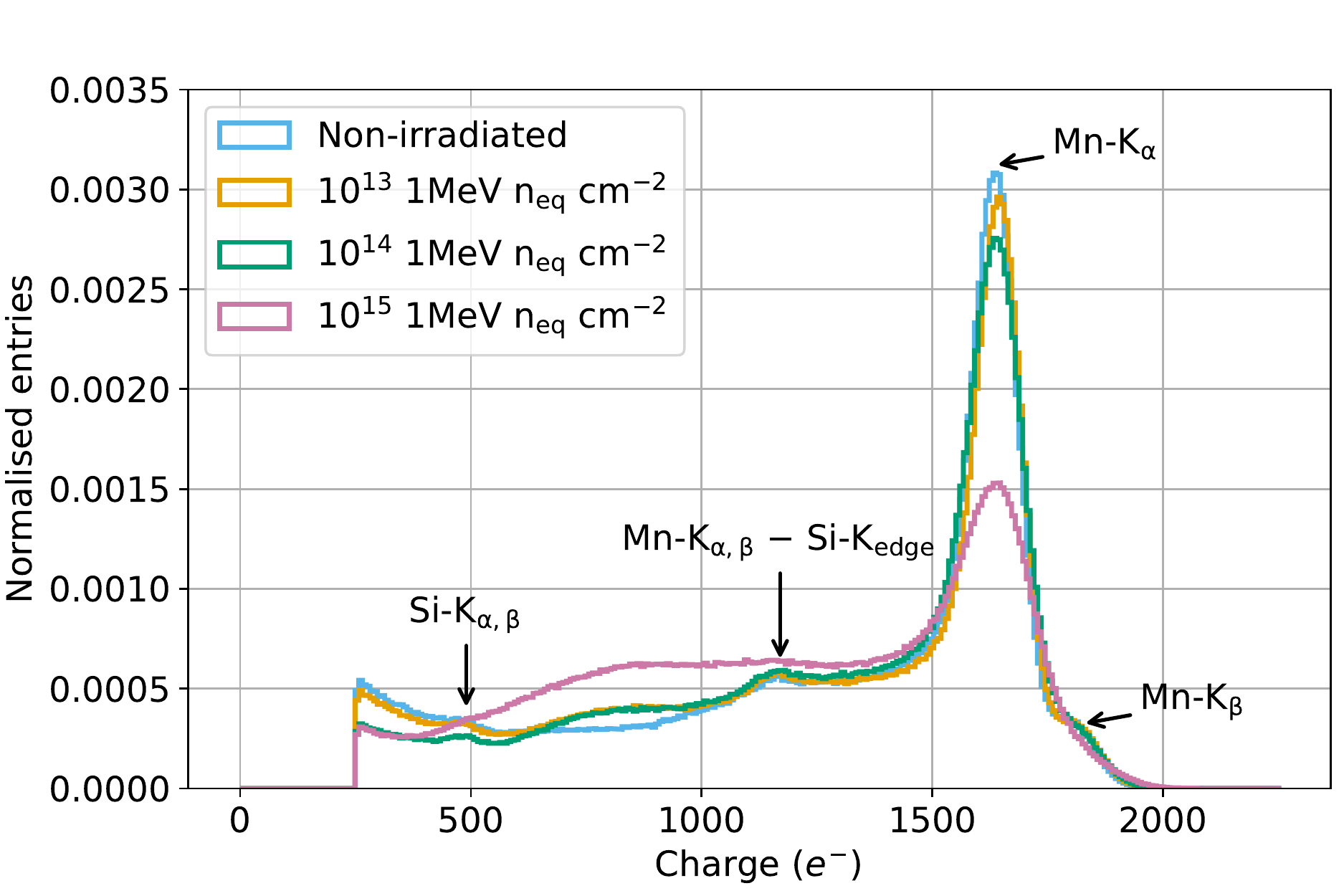}
    \caption{Measured $^{55}$Fe spectra of seed pixels for different levels of non-ionising irradiation: non-irradiated, \qtylist[list-units=single]{1e13;1e14;1e15}{\niel}.}
    \label{fig:fe55_irrad_comp}
\end{figure}

Figure~\ref{fig:fe55_irrad_comp} shows the measured $^{55}$Fe spectra for sensors irradiated with different non-ionising doses (non-irradiated, \qtylist[list-units=single]{1e13;1e14;1e15}{\niel}). For these measurements, the seed pixel ToT is plotted, where the seed pixel is defined as the pixel with the largest ToT in the set of adjacent pixels firing within the same oscilloscope capture. As the level of irradiation increases, the $\text{Mn-K}_{\alpha}$ peak becomes broader, with a notable broadening between \SI{1e14}{\niel} and \SI{1e15}{\niel}. 
These differences are shown in the resolution of the $\text{Mn-K}_{\alpha}$ peak between the irradiated sensors whose values are given in Table~\ref{tab:Fe55res}. The table shows that the largest increase in the resolution is from \SI{1e14}{\niel} to \SI{1e15}{\niel}. In addition, at the largest irradiation dose, the four peaks are no longer resolved and the contribution from seed pixels with energy in the range of \SIrange{400}{1400}{\ele} becomes more prominent. 
These changes to the spectrum at \SI{1e15}{\niel} indicate an alteration to the charge collection mechanisms in the sensor due to radiation damage, such as the increased recombination rate and the changes in the electric fields.

\begin{table}[!hbt]
\centering
\caption{Measured resolution of the $\text{Mn-K}_{\alpha}$ peak for  different levels of non-ionising irradiation in seed pixel signal spectra in Fig.~\ref{fig:fe55_irrad_comp}.}
\begin{tabular}{l c}
    \toprule
    Irradiation & Resolution (\si{\percent}) \\
    \midrule
    Non-irradiated & 7.40$\pm$0.02\\
    \SI{1e13}{\niel} & 7.42$\pm$0.03\\
    \SI{1e14}{\niel} & 8.73$\pm$0.02\\
    \SI{1e15}{\niel} & 12.05$\pm$0.07\\
    \bottomrule
\end{tabular}
\label{tab:Fe55res}
\end{table}
\section{Measurements with ionising particle beams}
The DPTS sensors were also characterised at facilities providing charged particle beams.
The results presented in this section are based on data taken from September 2021 to July 2022 at DESY~II \cite{desyII} and CERN~PS testbeam facilities. As such, the sensors were subject to normally incident \SI{5.4}{\giga\electronvolt\per c} electrons and  \SI{10}{\giga\electronvolt\per c} positive hadrons, respectively.

\subsection{Setup}
\label{sec:testbeam_setup}

\begin{figure}[!htb]
	\centering
    \input{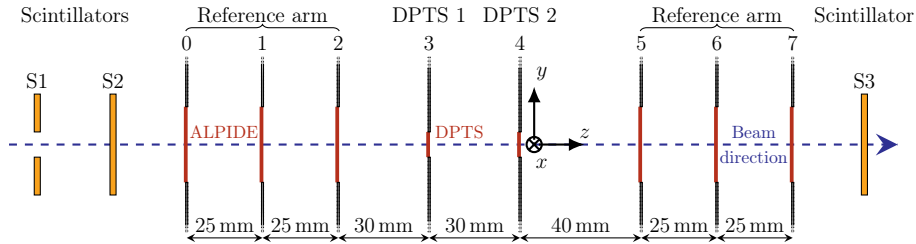}
    \caption{Sketch of the beam telescope (not to scale). Two DPTS are sandwiched between reference planes comprised of ALPIDE chips. Two scintillators (S2 and S3), operated in coincidence, and one featuring a \qty{1}{\mm} hole (S1), operated in anti-coincidence, are used for triggering. The trigger can also be provided by one of the two DPTS.}
    \label{fig:setup_sketch}
\end{figure}

A beam telescope comprising of reference planes equipped with ALPIDE chips \cite{ALPIDE-proceedings-1,ALPIDE-proceedings-2,ALPIDE-proceedings-3} has been used to reconstruct particle tracks. Two DPTS sensors were installed in-between the reference planes.
A sketch of the beam telescope is shown in Fig.~\ref{fig:setup_sketch}.
The data acquisition was based on the EUDAQ2 framework~\cite{eudaq}.
The trigger signal was provided either by one of the DPTS chips or by a discriminated output of photomultiplier tubes (operated in the plateau regime) connected to a set of three \qtyproduct{2x2}{\cm} scintillators placed in front and behind the telescope. One of the scintillators (S1) features a \qty{1}{\mm} hole in its centre and is operated in anti-coincidence with the other two ($\neg \text{S1} \wedge \text{S2} \wedge \text{S3}$,  cf.~Fig.~\ref{fig:setup_sketch}).
The $x$ and $y$ position of a DPTS and the anti-coincidence scintillator were adjustable via micro-positioning stages to achieve the overlap with the other DPTS.
The DPTS chip(s) not used for triggering are the device(s) under test or DUT(s).
An aluminum cooling jig (cf.~Sec.~\ref{sec:daq_setup}) with a \qty{1}{\mm} opening corresponding to the chip position was used to keep the DUT at a controlled temperature of \SI{+20}{\celsius} during the detection efficiency and position resolution measurements.

\subsection{Analysis tools and methods}
\label{sec:testbeam_analysis}

Data were processed in the Corryvreckan test beam reconstruction software framework~\cite{corryvreckan} by fitting General Broken Lines \cite{gbl_formalism} to clusters found in the reference planes and interpolating the tracks to the DUT(s).
Event and track quality selection criteria were applied to ensure a clean data sample: precisely one track per event, reduced track $\chi^2 < 3$, and track points on each reference plane.
Pixels on the reference planes with too large hit-rate (more than \num{1000}~times the average; \num{<1}~pixel per plane were affected) were masked.
Furthermore, the tracks intersecting the DUT within two pixel pitches from the sensor edge were rejected.

The DPTS hit pixel position was derived from the CML output corresponding to the assertion of the in-pixel discriminator (cf.~Sec.~\ref{sec:dpts:encoding}). The non-decodable events, resulting from the collisions on the CML lines when multiple pixels fire simultaneously, were associated to pixel $(15,15)$, i.e.\ to the centre of the matrix. Given the negligible likelihood that two fake hits coincide in time and result in a non-decodable event, these events are considered as real hits with an undetermined position. The fraction of such events was below \qty{3}{\percent}, for all chips and settings.

In the detection efficiency calculation, the DUT clusters were associated to tracks passing within a circular window with a radius of \qty{480}{\um}. The loose spatial cut prevents the underestimation of the efficiency that would result from the exclusion of non-decodable events associated to pixel $(15,15)$.
In order to minimise the probability of associating a fake (noise) hit to a track, the DUT response was required to be within \SI{1.5}{\micro\second} of the trigger signal. 
The efficiency of the DUT is then estimated by the fraction of tracks with associated clusters. The relative uncertainties are obtained by calculating the \qty{68.3}{\percent} Clopper-Pearson confidence interval and summing in quadrature the probability of associating a fake hit (given by the in-situ measured fake-hit rate, cf.~Sec.~\ref{sec:lab_meas_fhr} and Sec.~\ref{sec:testbeam_results_eff}).

For the position resolution evaluation, a spatial acceptance window with a radius of \qty{45}{\um} around the track and a time acceptance window of \SI{1.5}{\micro\second} were applied. The distance between the intercept on the DUT plane and the associated cluster position (given by the centre of mass of pixels in the cluster) in column and row direction is determined for each track, yielding two spatial residual distributions. By fitting the latter with two Gaussian functions, the standard deviation parameters $\sigma_{col}$ and $\sigma_{row}$ are obtained. 
The position resolution is then retrieved by quadratically subtracting the estimated telescope tracking resolution of $\sigma_{track}=\qty{2.4}{\um}$ \cite{telopt}.
Given that the obtained position resolution in the two directions is equal within the measurement precision and compatible with square pixel geometry, their average will be referred to as the position resolution in the rest of the paper.

The timing resolution analysis involved no time acceptance window, while the spatial acceptance window was the same one used in the position resolution analysis.

\subsection{Detection efficiency and fake-hit rate}
\label{sec:testbeam_results_eff}

\begin{figure}[!thb]
	\centering
	\begin{subfigure}[t]{0.99\textwidth}
    \includegraphics[width=\textwidth]{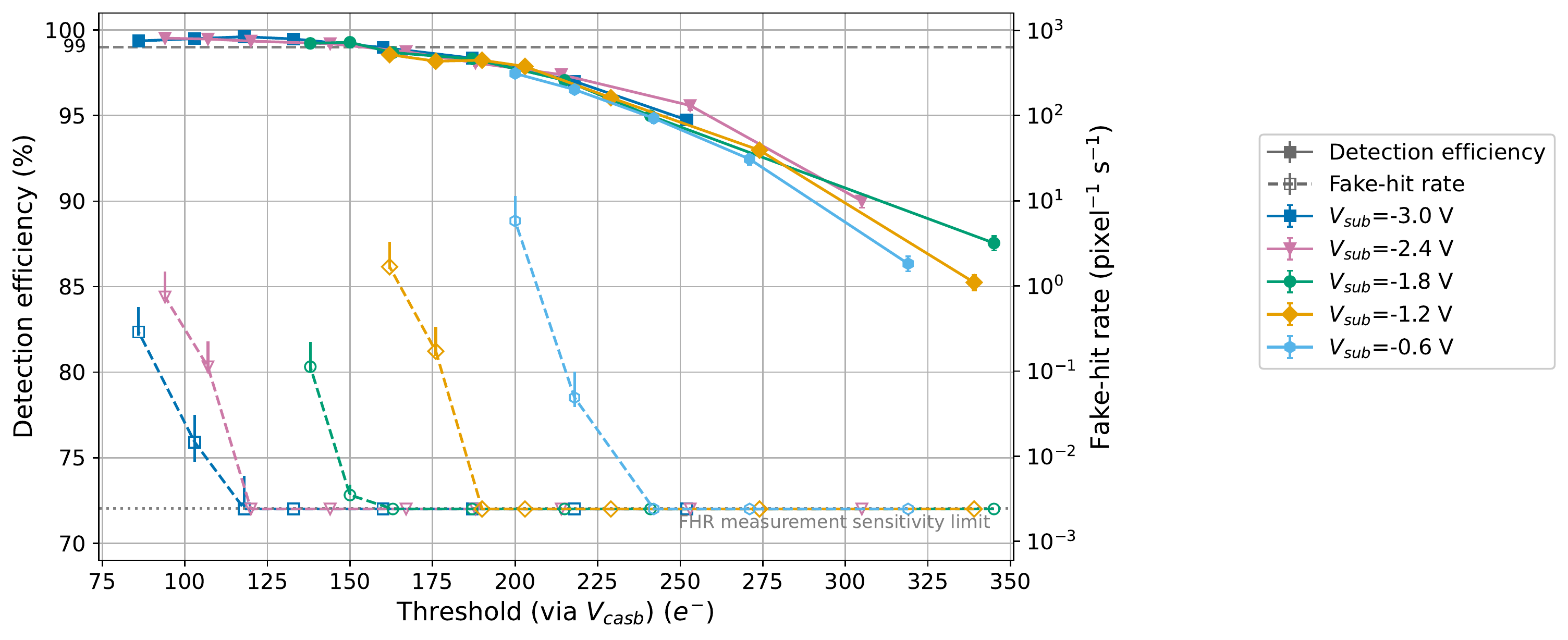}
    \caption{Sensor irradiated to a dose of \SI{10}{\kilo\gray} and \SI{e13}{\niel} at different $\Vsub=\Vpwell$.
    }
    \label{fig:2022-07_PS_B3_efficiency}
    \vspace{0.3cm}
	\end{subfigure}
	\begin{subfigure}[t]{0.95\textwidth}
    \includegraphics[width=\textwidth]{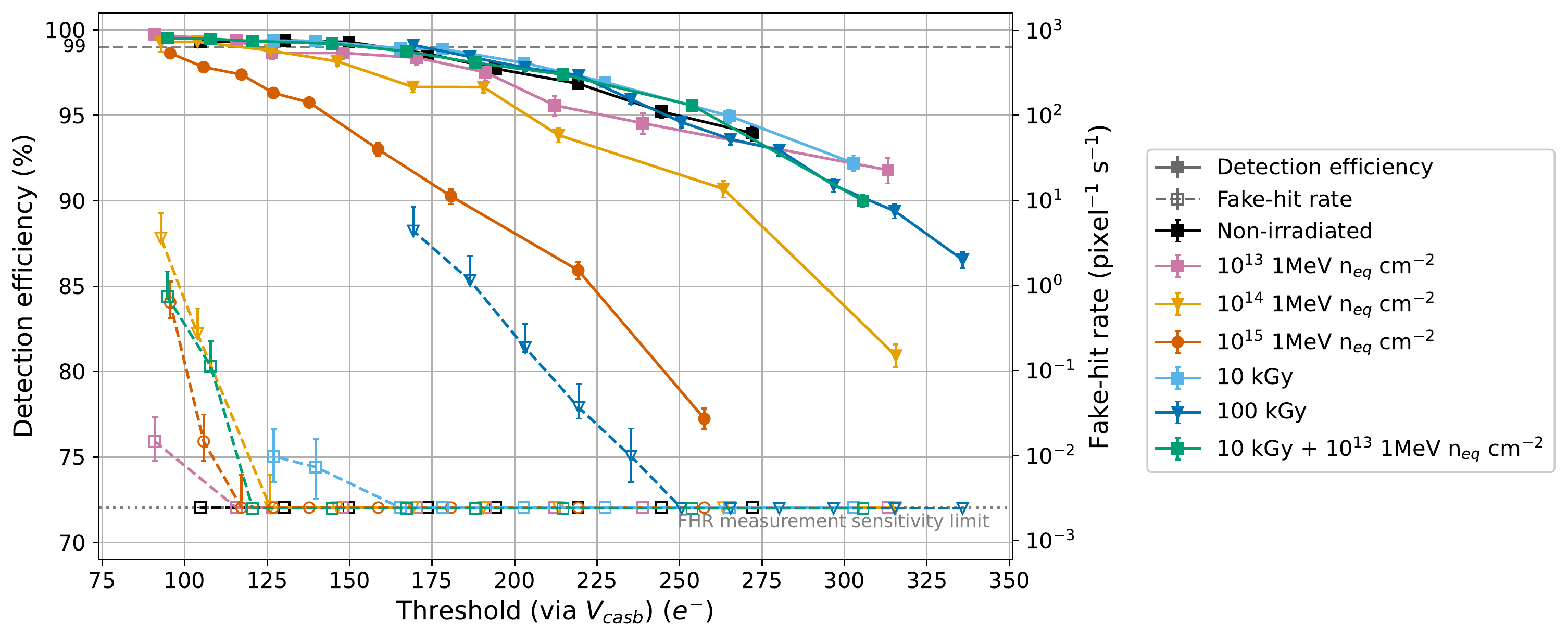}
    \caption{Sensors irradiated to different levels.
    }
    \label{fig:2022-07_PS_efficiency_irradiation_comparison}
	\end{subfigure}
    \caption{Detection efficiency (filled symbols, solid lines) and fake-hit rate (open symbols, dashed lines) as a function of average threshold, measured with \qty{10}{\giga\electronvolt\per c} positive hadrons.}
\end{figure}

Figure~\ref{fig:2022-07_PS_B3_efficiency} shows the detection efficiency and the fake-hit rate (measured in situ) as a function of the average chip threshold set by changing $\Vb$ at different reverse biases for a sensor irradiated with protons, i.e.\ that has received a combination of an ionising dose of \SI{10}{\kilo\gray} and a non-ionising dose of \SI{e13}{\niel} (levels compatible with the ITS3 requirements).
Instead of showing the data for a non-irradiated sensor, which exhibits fake-hit rates below the measurement sensitivity limit, this particular sensor was chosen to better illustrate the effect of the reverse bias on the performance. In particular, it can be observed that by increasing $\Vsub=\Vpwell$, the onset of the measured fake-hit rate is offset to lower thresholds, thus increasing the operational margin at above \qty{99}{\percent} detection efficiency.

The effect of different irradiation levels on the detection efficiency and the fake-hit rate is shown in Fig.~\ref{fig:2022-07_PS_efficiency_irradiation_comparison}, where the reverse bias is kept at \SI{-2.4}{\volt}.
It can be observed that non-ionising irradiation leads to a decrease in the detection efficiency, while ionising irradiation leads to an increase in the fake-hit rate.
These trends are consistent with the expectation that the largest effect of the non-ionising and ionising radiation damage is on the charge collection and the noise (front-end) performance, respectively.
For the \SI{100}{\kilo\gray} irradiated sensor, a significant increase in the fake-hit rate can be noticed, with onset at a much higher threshold than in the other cases.
Finally, although the \SI{e15}{\niel} irradiated sensor shows notable performance deterioration, it can still be operated at \qty{99}{\percent} efficiency at the temperature of \qty{+20}{\celsius}.

\begin{figure}[!htb]
	\centering
    \includegraphics[width=0.99\textwidth]{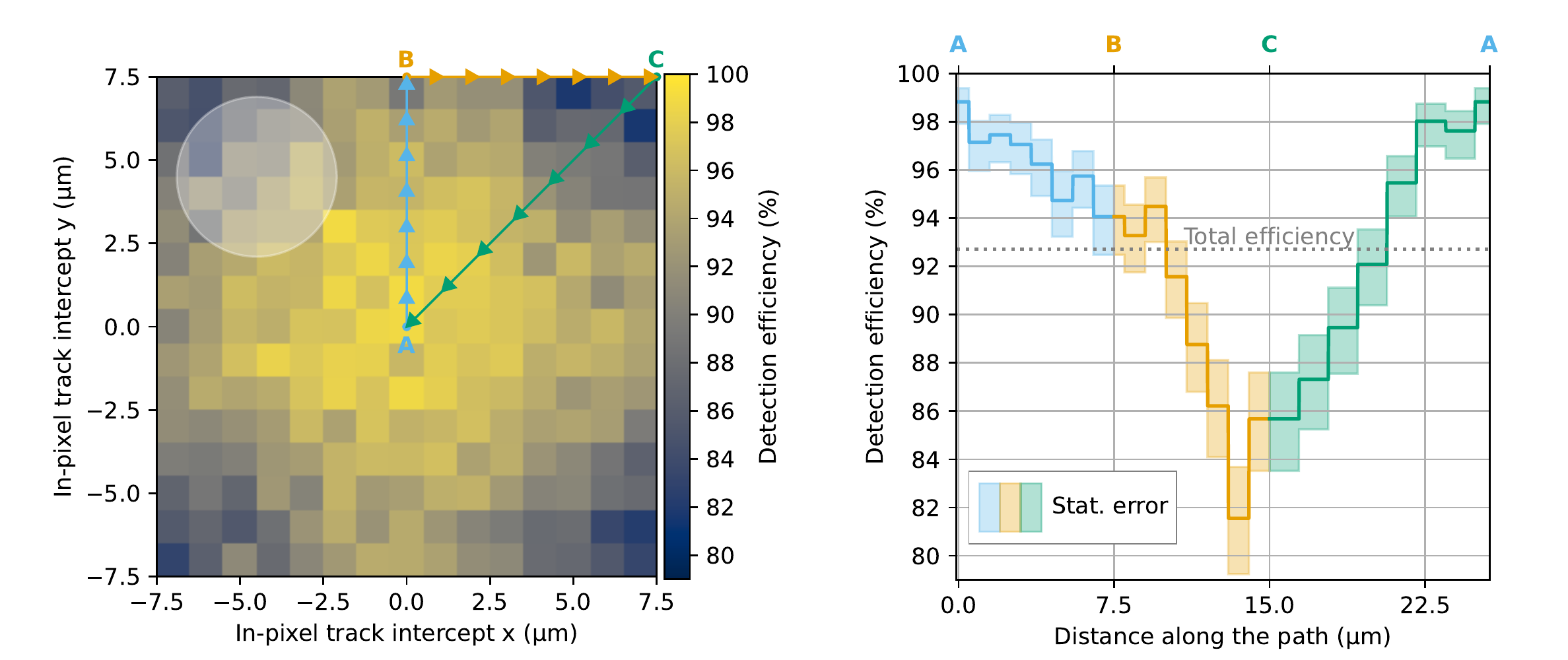}
    \caption{In-pixel detection efficiency of a \qty{e15}{\niel} irradiated sensor with threshold set to \qty{160}{\ele}, measured with \qty{10}{\giga\electronvolt\per c} positive hadrons. The tracking resolution is represented by the white circle ($r=\sigma_{track}=\qty{2.4}{\um}$) in the top left corner.
    }
    \label{fig:inpix_eff}
\end{figure}

The origin of the detection efficiency loss was investigated by studying its dependency on the particle hit position within a pixel. Figure~\ref{fig:inpix_eff} shows the detection efficiency of a \qty{e15}{\niel} irradiated sensor as a function of reconstructed track position relative to the nearest pixel centre.
As expected, the detection efficiency decreases when the particle track is further away from the collection diode. A similar result was obtained for a non-irradiated sensor, with the same magnitude of efficiency loss observed at higher thresholds.

\subsection{Spatial resolution and cluster size}
\label{sec:testbeam_results_resolution}

\begin{figure}[!th]
    \centering
	\begin{subfigure}[t]{0.95\textwidth}
    \includegraphics[width=0.99\textwidth]{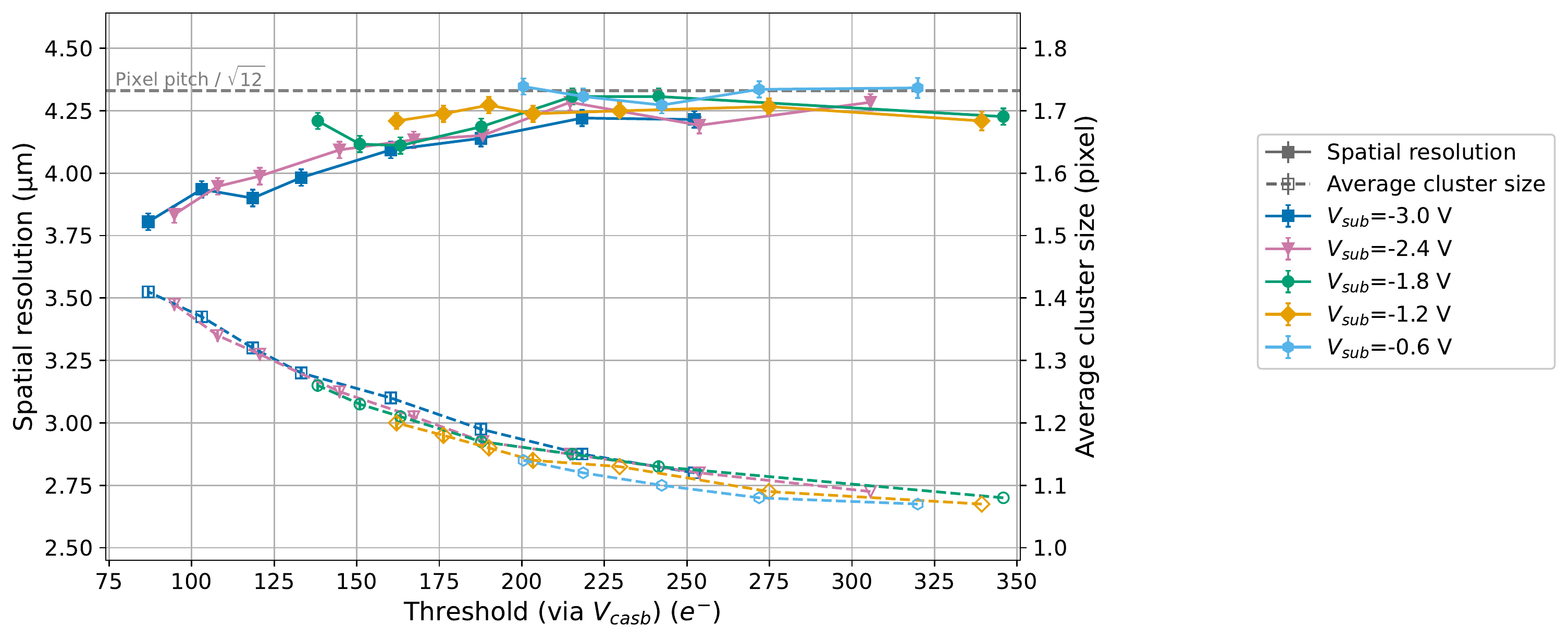}
    \caption{Sensor irradiated to a dose of \SI{10}{\kilo\gray} and \SI{e13}{\niel} at different $\Vsub=\Vpwell$.
    }
    \label{fig:2022-07_PS_B3_spatial_resolution}
    \vspace{0.3cm}
    \end{subfigure}

	\begin{subfigure}[t]{0.99\textwidth}
    \includegraphics[width=0.99\textwidth]{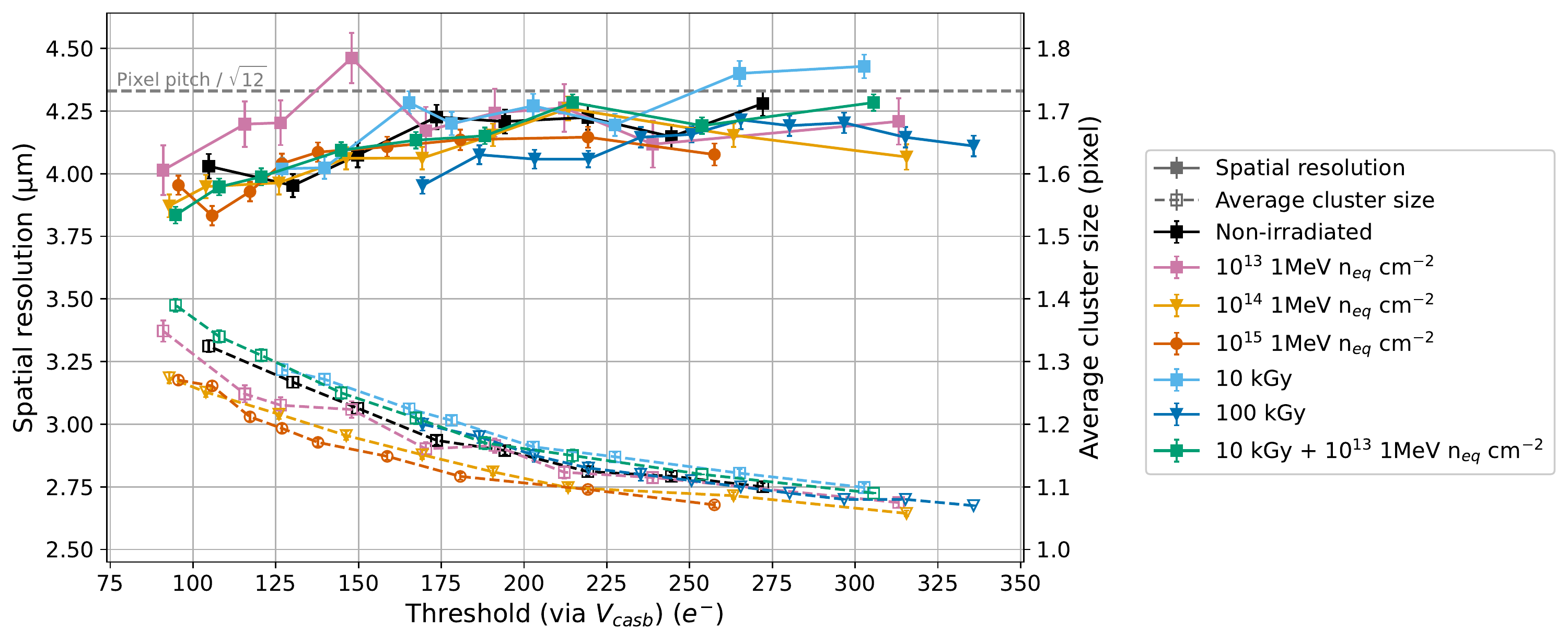}
    \caption{Sensors irradiated to different levels.
    }
    \label{fig:2022-07_PS_different_irradiation_levels_spatial_resolution}
    \end{subfigure}
    \caption{Spatial resolution (solid lines) and average cluster size (dashed lines) as a function of the average threshold, measured with \qty{10}{\giga\electronvolt\per c} positive hadrons.}
 \end{figure}

Figure~\ref{fig:2022-07_PS_B3_spatial_resolution} shows the sensor spatial resolution and average cluster size as a function of the average threshold for different $\Vsub$ applied, measured with \qty{10}{\giga\electronvolt\per c} hadrons for the device irradiated with both an ionising radiation dose of \SI{10}{\kilo\gray} and a non-ionising radiation dose of \SI{e13}{\niel} (levels compatible with the ITS3 requirements).
The benefit of applying higher $\Vsub$ is not evident without combining the information from Fig.~\ref{fig:2022-07_PS_B3_efficiency}, which shows that higher $\Vsub$ is necessary to operate at lower thresholds. In this regime, the additional positional information, given by a larger average cluster size, results in a spatial resolution better than that of a purely binary sensor (pixel pitch divided by $\sqrt{12}$).

The effect of different irradiation levels on the spatial resolution and the average cluster size is demonstrated in Fig.~\ref{fig:2022-07_PS_different_irradiation_levels_spatial_resolution}, where the reverse bias is kept constant at \SI{-2.4}{\volt}.
Regardless of the irradiation level, the measured spatial resolution is equal or even slightly better than pixel pitch divided by $\sqrt{12}$, with no degradation of the spatial resolution performance related to the received dose.
The average cluster size exhibits a slight, but systematic, decrease with the increasing non-ionising radiation dose. This trend is compatible with the previous observations of a deteriorated charge collection process.

\subsection{Timing resolution}
\label{sec:testbeam_results_time}

The timing resolution was estimated by measuring the output signal coincidence of two DPTS using a \SI{5.4}{\giga\electronvolt\per c} electron beam. The trigger signal and thus the time reference is given by the scintillators signal (cf.~Fig.~\ref{fig:setup_sketch}).

\begin{figure}[!th]
    \centering
    \includegraphics[width=0.9\textwidth]{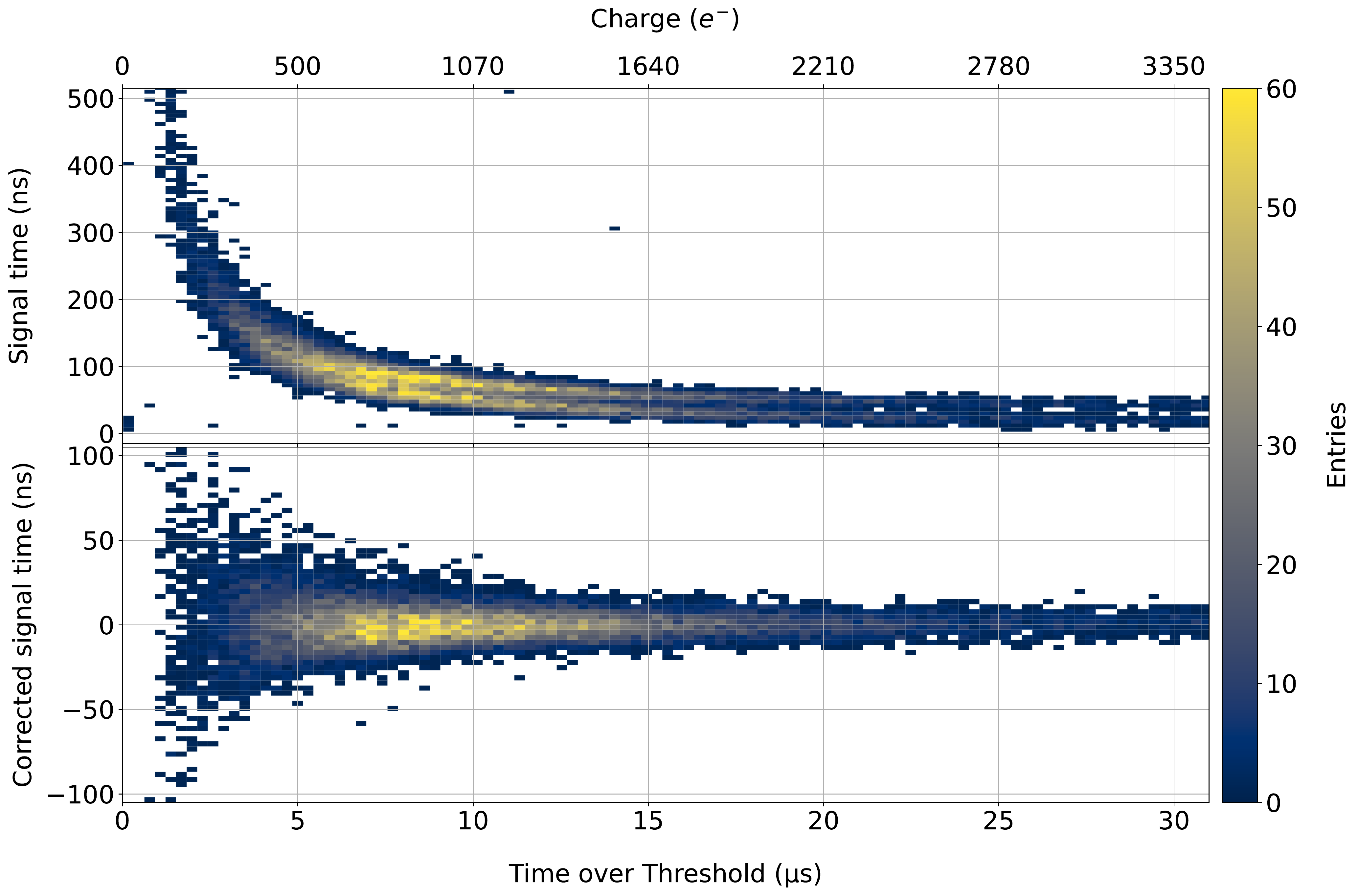}
    \caption{Signal time vs ToT for the upstream DPTS measured with \SI{5.4}{\giga\electronvolt\per c} electrons, with no corrections applied (top) and after readout scheme and time walk corrections (bottom). The conversion to charge is obtained by applying the procedure discussed in Sec.~\ref{sec:lab_meas_55Fe}.
    }
    \label{fig:2021-09_DESY_DPTSOW22B3_ToTvsToA_correlation}
\end{figure}

Figure~\ref{fig:2021-09_DESY_DPTSOW22B3_ToTvsToA_correlation} (top), shows the correlation of the signal arrival time and ToT, i.e.\ signal amplitude, for one of the two DPTS chips. 
The influence of the chip readout scheme on the signal time response is notable at high ToT, where the distribution splits into two tails which correspond to odd and even columns (cf.~Sec.~\ref{sec:dpts:encoding}). The correction for this effect is applied by subtracting the asymptotic value of the two tails from even and odd columns, respectively.
The time walk observed for the lower input amplitude is corrected by fitting the data with the empirical function

\begin{equation}
    \label{eq:ToT_fit}
    \text{Signal time} = A + \frac{B}{ToT - C}\text{ ,}
\end{equation}

\noindent where $A$, $B$ and $C$ are fit parameters, and then subtracting its value from the measured data points.
The signal (arrival) time vs ToT correlation corrected for the chip readout scheme and the time walk is shown in Fig.~\ref{fig:2021-09_DESY_DPTSOW22B3_ToTvsToA_correlation} (bottom).
As intended, it can be observed that the amplitude of the time walk is reduced by a factor of about \num{3} and its asymmetric feature is removed. The correlation between the amplitude of the time walk and the time-over-threshold remains but was not further investigated in this study.

\begin{figure}[!th]
    \centering
    \includegraphics[width=0.9\textwidth]{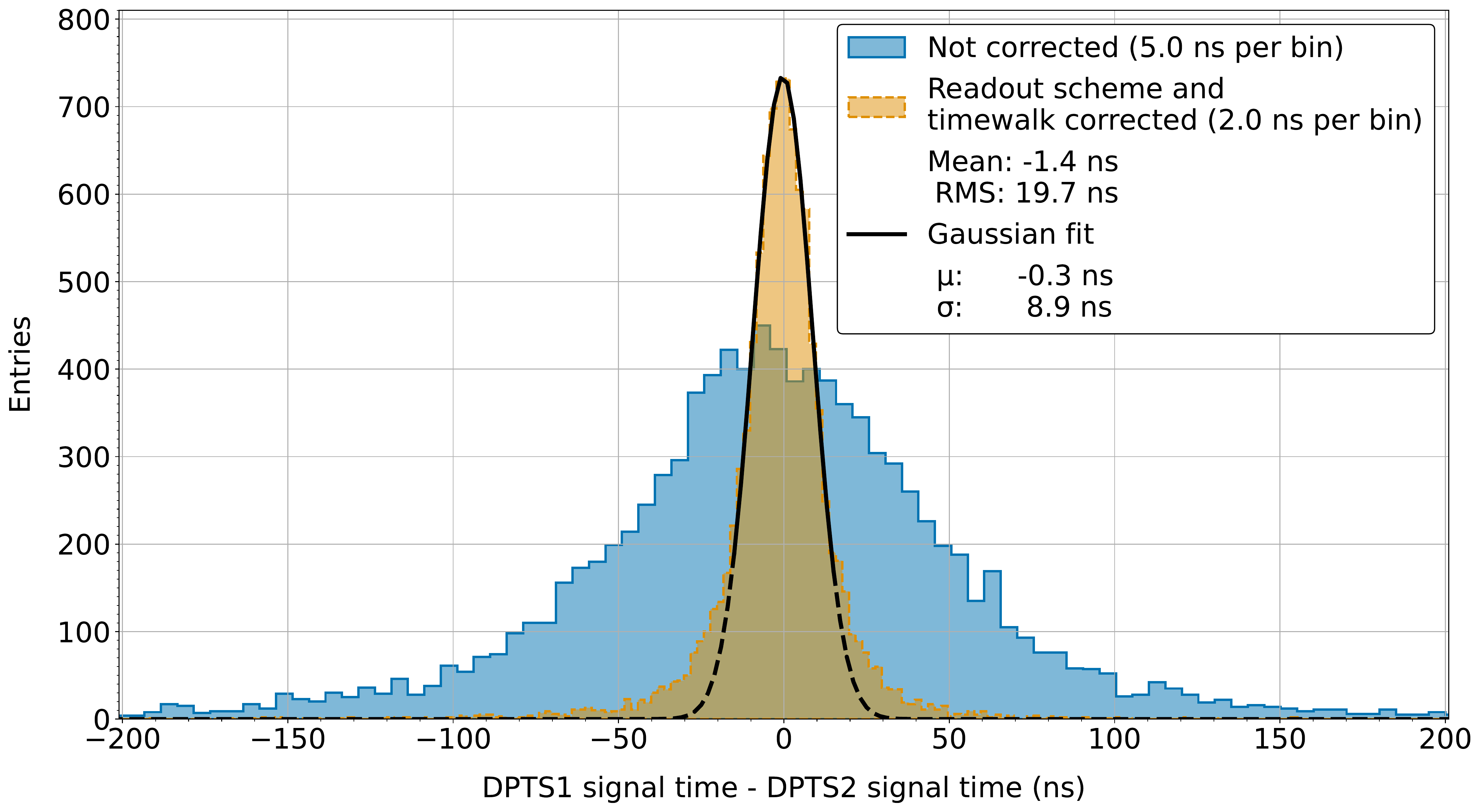}
    \caption{Time residuals distributions of two DPTSs measured with \SI{5.4}{\giga\electronvolt\per c} electrons with no corrections (blue) and with readout scheme and time walk corrections applied (orange). The corrected distribution is fitted with a Gaussian function in the time residuals range from $\qty{-15}{ns}$ to $\qty{15}{ns}$ (black solid line, dashed line for points outside the fit range).
    }
    \label{fig:2021-09_DESY_DPTS_time_residuals}
\end{figure}

The distribution of time residuals, i.e.\ the time differences of the output of the two DPTSs is shown in Fig.~\ref{fig:2021-09_DESY_DPTS_time_residuals} for non-corrected data (in blue) and with corrections discussed above applied independently for the two DPTSs (in orange).
The timing resolution of a single DPTS is obtained by fitting the distribution with a Gaussian, and the fit parameter $\sigma$ is divided by $\sqrt{2}$, resulting in $\sigma_{t} = \qty{6.3}{ns} \pm \qty{0.1}{ns}$~(stat).
It is worth highlighting that this result is obtained by operating the chips at the low end of the in-pixel front-end power consumption range (i.e.\ nominal conditions in Sec.~\ref{sec:daq_setup}). An improvement in timing performances is expected by increasing the front-end current, namely $\Ib$ and $\Ibn$.

\section{Conclusions}
Laboratory and in-beam measurements have been used to systematically validate the charge detection and radiation hardness of the Digital Pixel Test Structure MAPS prototype submitted in the TPSCo~\SI{65}{\nm} process. 
The front end has shown a robust performance that can maintain a suitable operation regime in terms of threshold spread, noise and fake-hit rate. In addition to measuring the ToT response from X-ray emissions of an $^{55}$Fe source, the DPTS achieved a resolution of $\qty{7.40}{\percent}\pm\qty{0.02}{\percent}$ for the $\text{Mn-K}_{\alpha}$ peak. The impact of irradiation on the performance becomes evident at \SI{1e15}{\niel}; however, the sensor is still able to effectively detect the X-ray emissions. This demonstration of the radiation hardness is reinforced further by the excellent detection efficiency of \SI{99}{\percent} and spatial resolution below the binary resolution for chip irradiation doses compatible with the ITS3 requirements (\qty{e13}{\niel} and \qty{10}{\kilo\gray}) as well as to doses above these levels at \SI{1e15}{\niel} and \SI{100}{\kilo\gray} at +\SI{20}{\celsius} while preserving a fake-hit rate below \num{10}~pixel$^{-1}$~s$^{-1}$. The DPTS also demonstrates a timing resolution of about \SI{6}{\ns} for the nominal chip bias settings, with improvements expected if operated at conditions optimised for timing performance.

The first step to validate this sensor technology has been presented and makes up an important aspect of the R\&D for the ALICE ITS3 as well as a significant contribution to the CERN EP R\&D on monolithic sensors. The excellent performance of the DPTS opens the way for further developments in this technology and with this sensor design. The next step towards a wafer-scale bent sensor and a fully cylindrical detector is the validation of stitching and yield via the full-scale sensor prototypes produced in the second submission in the \qty{65}{\nm} process, designated ER1.

Further work to investigate the limit of the radiation hardness of this technology and probe the performance of the DPTS with ionising particles at non-zero incident angles is foreseen.
Moreover, the front-end operating point is being tuned to optimise the power consumption for low power applications, such as ALICE ITS3, where high radiation hardness and fast timing response are not crucial requirements.

\newenvironment{acknowledgement}{\relax}{\relax}
\begin{acknowledgement}
\section*{Acknowledgements}
The measurements leading to these results have been performed at the Test Beam Facility at DESY Hamburg (Germany), a member of the Helmholtz Association (HGF). We would like to thank the coordinators at DESY for their valuable support of this testbeam measurement and for the excellent testbeam environment.

P.~Becht acknowledges the support by the HighRR research training group [GRK~2058] and funding by the German Federal Ministry of Education and Research/BMBF (project number 05H21VHRD1) within the High-D consortium.

W.~Deng acknowledges the support by the National Key Research and Development Program of China (2022YFA1602103).

This work has been sponsored by the Wolfgang Gentner Programme of the German Federal Ministry of Education and Research (grant no.~13E18CHA).    
\end{acknowledgement}
%

\bibliography{references}

\end{document}